\documentclass[12pt]{article}
\usepackage{color,epsfig,amsmath,amssymb,graphicx}

\def\be{\begin{equation}}
\def\fe{\end{equation}}
\def\spose#1{\hbox to 0pt{#1\hss}}
\def\lta{\mathrel{\spose{\lower 3pt\hbox
{$\mathchar"218$}}\raise 2.0pt\hbox{$\mathchar"13C$}}}  \def\gta{\mathrel
{\spose{\lower 3pt\hbox{$\mathchar"218$}}\raise 2.0pt\hbox{$\mathchar"13E$}}}

\def\Libra{\spose {--} {\cal L}}
\def\Euro{\spose {\lower 2.5pt\hbox{${^{\bf =}}$}}{ C}}

\def\spose#1{\hbox to 0pt{#1\hss}}

\def\sqr#1#2{{\vcenter{\hrule height.4pt\hbox{\vrule width.8pt height#2pt
\kern#1pt\vrule width.8pt}\hrule height.4pt}}}

\definecolor{violet}{rgb}{0.4,0,0.4}
\definecolor{vert}{rgb}{0,0.5,0.0}
\definecolor{navy}{rgb}{0.0,0.0,0.6}
\definecolor{orange}{rgb}{0.8,0.2,0.0}
\definecolor{bleu}{rgb}{0.3,0.0,0.8}
\definecolor{brun}{rgb}{0.6,0.3,0.0}

\def\FF{ {\color{vert} F}}
\def\BB{ {\color{vert} B}}
\def\AA{ {\color{vert} A}}

\def\Ff{ {\color{black} F}}
\def\Bb{ {\color{black} B}}
\def\Aa{ {\color{black} A}}

\def\rms{ {\rm S}} \def\rmf{ {\rm f}} 
  \def\rmc{ {\rm c}}
\def\Armc{ {{\color{vert}\rm A}_\rmc}}
\def\rmI{ {\rm I}} 

\def\nn{{\color{vert}n}}
\def\qq{{\color{vert}q}}
\def\nc{\nn_{\rmc}}
\def\nf{\nn_{\rmf}}

\def\EU{{\color{brun}\Euro}}
\def\rrho{{\color{brun}\rho}}
\def\TT{{\color{brun}T}}

\def\ff{{\color{brun}f}}
\def\lamb{{\color{brun}{\lambda}}}
\def\LP{{\color{brun}{\cal P}}}
\def\PP{{\color{brun}{P}}}
\def\QQ{{\color{brun}{Q}}}

\def\Shear{{\color{brun}{\mathit \Sigma}}}
\def\LLambda{{\color{brun}\Lambda}}
\def\PPsi{{\color{brun}\Psi}}

\def\LmP{{\color{red}{\cal P}}}
\def\calE{{\color{red}{\cal E}}} 
\def\mmm{{\color{red}m}}
\def\mm{{\mmm_{\rm o}}}

\def\hhbar{{\color{red}\hbar}}

\def\ww{{\color{red} w}}
\def\pp{{\color{red}{p}}}

\def\hhbar{{\color{red}{\hbar}}}

\def\uu{{\color{violet}u}} \def\vv{{\color{violet}v}}

\def\cc{{\color{bleu}c}}

\def\xx{{\color{bleu}x}} 
\def\gr{{\color{bleu}g}}
\def\ggamma{{\color{bleu}\gamma}}

\def\ddelta{{\color{bleu}\delta}}

\def\gamm{{\color{violet}\gamma}}
\def\epsil{{\color{violet}\epsilon}}
\def\sigm{{\color{violet}\varsigma}}

\def\varpphi{{\color{violet}\varphi}}

\def\ppi{{\color{red}\pi}}

\def\mmu{{\color{red}\mu}} 
 
\def\cchi{{\color{red}\chi}}

\begin{document}

\title{\color{violet}\bf Relativistic  mechanics of neutron superfluid in 
(magneto) elastic star crust.}

\author {{\bf \color{brun}  Brandon Carter}$^\star$
{\bf \color{brun} and Lars Samuelsson}$^\dagger$
\hskip 1 cm\\
\\ $^\star$ {\color{vert} LUTh, Obs. de Paris, 92195 Meudon, France;}\\
(Brandon.Carter@obspm.fr)\\
 $^\dagger${\color{vert} Mathematics, Univ. Southampton, SO17 IBJ, U.K.}\\
(lars@soton.ac.uk)}
\date{\color{bleu}\it May, 2006}

\maketitle

\vskip 1.6  cm

{\bf Abstract}  At densities below the neutron drip threshold, a 
purely elastic solid model (including, if necessary, a frozen-in
magnetic field) can provide an adequate description of a neutron 
star crust, but at higher densities it will be necessary to allow 
for the penetration of the solid lattice by an independently moving 
current of superfluid neutrons. In order to do this, the previously 
available category of relativistic elasticity models is combined 
here with a separately developed category of relativistic 
superfluidity models in a unified treatment based on the use of an 
appropriate Lagrangian master function. As well as models of the 
purely variational kind, in which the vortices flow freely with the 
fluid, such a master function also provides a corresponding
category of non-dissipative models in which the vortices are pinned 
to the solid structure.

\vfill\eject
\section{Introduction}

The purpose of this work is to adapt the previously developed
formalism for the treatment of various kinds of conductivity in a
relativistic solid \cite{C89} to the particular case of a neutron
superfluid in the solid crust of a neutron star. This treatment will
include allowance for the presence of a frozen-in magnetic field in
the limit of perfect electrical conductivity, which will be a very
good approximation in this context. (Such an M.H.D. type limit is more
relevant in neutron stars than the kind of diamagnetic polarisation
that was considered in earlier work \cite{M78,C80}.) The effects of
magnetic fields may be significant for ordinary pulsars and will be
particularly important in the special case of magnetars. In fact, the
recently observerd quasi periodic oscillations in the aftermath of
giant flares in the soft gamma ray repeaters SGR 1806-20
\cite{IETAL05,WS06} and SGR 1900+14 \cite{SW05} are suggested to be
associated with oscillations originating in the crust after a major
quake, see \cite{D98,SW05,GSA06} and references therein. The magnetic
field will be of crucial importance for a complete understanding of
such oscillations.

One of the main motivations for this work is to deal with the
frequency glitches in pulsars, whose treatment has long been
recognised to require allowance both for elastic solidity and for the
independent superfluid motion of neutrons in the crust at densities
above the ``drip'' threshold (at about $10^{11}$g/cm$^3$). In the
literature on relativistic continuum mechanics, the superfluid aspects
(whose development is relatively recent) have mainly been treated
separately \cite{CK92,Langlois98,ACL02,Comer04,PNC05} from the purely
elastic aspects, for which the formalism previously developed for that
purpose \cite{CQ72,C73,FS75,CQ77} has already been applied
\cite{CQ05,Q76,ST83,F90,P92,KS02,KS03}  to fairly realistic kinds of 
neutron star models (as well as to idealised configurations 
\cite{Beig02,KS04} of more artificial kinds that are of essentially 
mathematical rather than astrophysical interest). A synthesis of the kind 
that is needed \cite{Cham06} for the treatment of neutron conduction 
through the solid crust has however been recently developed in a Newtonian 
framework \cite{CC06}, which should be accurate enough for the treatment 
at a local level of the outer and middle layers of a neutron star crust. 
Nevertheless, for the global treatment of a neutron star, and even for a 
local treatment of the deeper layers, a quantitatively precise 
description will require a fully relativistic treatment.

As a step toward the achievement of this purpose, the present work
will develop a treatment that is general relativistic in the sense of
being applicable to a space time background with an unrestricted
Lorentzian metric with signature $\{-,+,+,+\}$ having components
$\gr_{\mu\nu}$ with respect to arbitrary space-time coordinates
$\xx^\mu$. This metric will not be restricted by any specific
evolution laws, so the treatment will be compatible not only with the
standard Einstein theory of gravity but also with conceivable
alternatives, and in particular with the trivial flat background case
of special relativity, which will always be useful as an approximation
in a local neighbourhood. We wish to stress that the formalism is not
restricted to fixed backgound spacetimes. In particular, as an
example, it is applicable to the full dynamical Einstein equations,
for which it provides the right hand side.

The Newtonian treatment referred to above \cite{CC06} included allowance 
for dissipative effects such as vortex creep, but -- to avoid introducing 
too many complications at once -- the relativistic treatment provided 
here will be of conservative type, applying to the low  temperature limit 
in which dissipative effects are neglected. As a generalisation of the 
variational formulation \cite{C73} of the purely elastic relativistic 
solid model \cite{CQ72} that is adequate for treating the outer crust 
layers (at densities below the ``drip'' threshold) the present treatment 
will be based on the use of an action density $\tilde\Lambda$ of a kind
that can be used as a Lagrangian to obtain models of strictly variational 
type, but that can also be used (in the manner to described in Subsection 
\ref{pineq} ) as a master function for the construction of non dissipative
models of more general type such as are capable, as has already been shown
in the relevant multiconstituent fluid limit case \cite{Langlois98} of 
incorporating the potentially important effect of vortex pinning. 

The category of conducting solid models developed here is designed for 
application on a macroscopic scale that is large compared with the 
intervortex separation. However as a special zero vorticity case, 
applicable on a mesoscopic scale -- large compared with the microscopic 
scale of internuclear spacing, but small compared with the intervortex 
separation  -- this category includes models of the relatively simple 
hyperelastic subcategory \cite{C06} that is also of interest in a 
cosmological context.

The requisite Lagrangian master function $\tilde\Lambda$ will be formulated 
in terms of a solid structure, representing a distribution of atomic nuclei 
in terms of a 3-dimensional material base manifold with position coordinates 
$\qq^{\rm_A}$, together with a current $\nn_\rmf^{\,\mu}$ of free baryons, 
representing superfluid neutrons, and a current  $\nn_\rmc^{\,\mu}$ of 
confined baryons, representing the protons and the fraction of the neutrons 
that are confined within the nuclei, so that their flow will be restricted 
by the comovement condition that is expressible (using a comma for partial 
derivation) by
{\be  \nn_\rmc^{\,\nu}\,\qq^{\,\rm_A}_{\, ,\nu}=0\, .\label{(2} \fe}

\section{Relativistic action formulation}

\subsection{Dynamic variables and background fields}

As well as depending on the ``live'' dynamical fields, consisting in this 
case of the (vectorial) current components $\nn_\rmf^{\,\mu}$ and 
$\nn_\rmc^{\,\mu}$ and the (scalar) base space coordinates $\qq^{\rm_A}$ 
and their derivatives, the Lagrangian will also depend on the metric 
$\gr_{\mu\nu}$ which in the present treatment will be considered as a
``dead'' -- meaning predetermined -- background. In such a case the most 
general infinitesimal Eulerian (i.e.\ fixed point) variation of the  
Lagrangian will be decomposable as the sum of two contributions in the 
form
{\be \delta\tilde\LLambda=\delta^{_\heartsuit\!}\tilde\LLambda +
\delta^\ddagger\tilde\LLambda\label{(0)} \fe}
in which $\delta^{_\heartsuit\!}\LLambda$ is the realisable part 
attributable to a physically possible alteration of the configuration of 
the ``live'' dynamical fields, while $\delta^\ddagger\LLambda$ is a virtual 
part arising from mathematically conceivable but (in the context under 
consideration) physically forbidden variations of the ``dead'' background 
fields that have been fixed in advance, of which the only one in the 
present instance is the flat or curved spacetime  metric $\gr_{\mu\nu}$,
so that the complete background variation contribution will be given just 
by 
{\be  \delta^\ddagger\tilde\LLambda=\frac{\partial\tilde\LLambda}
{\partial\gr_{\mu\nu}}\delta\gr_{\mu\nu}\, .\label{(1)}\fe}
The particular kind of background variation to be considered in this 
section is one that is simply generated by the action of a displacement 
field $\xi^\nu$ say, so that the ensuing field variations will be given 
just by the negatives of the corresponding Lie derivatives. For the 
background metric the resulting variation will be given (using round 
brackets indicate index symmetrisation)  by an expression of the familiar 
form
{\be \delta\gr_{\mu\nu}=-2\nabla_{\!(\mu}\xi_{\nu)}\, ,\label{(3)}\fe}
in which $\nabla_{\!\mu}$ denotes the (components of the) Riemannian 
covariant differentiation operator 
%the operator terms of Riemannian 
%covariant differentiation 
with respect to $\gr_{\mu\nu}$.

In so far as the ``live'' contribution is concerned, a further subdivision
arises in cases such as those of interest here, which are characterised
by a constrained variation principle, meaning one whereby the ``on shell'' 
evolution condition is that of invariance of the relevant action integral 
with respect to a compactly supported perturbation of the dynamical fields 
that is not entirely arbitrary but constrained by an appropriate 
admissibility condition. In the applications under consideration, the 
relevant admissibility condition will be interpretable as the requirement 
for the change to represent a ``natural'' variation of the same given 
physical system, whereas more general changes would represent a replacement
of the system by a slightly different system within the same category.
In such a case the ``live'' field contribution will be decomposable (though 
not necessarily in a unique manner) as a sum of the form
{\be \delta^{_\heartsuit\!}\tilde\LLambda=\delta^\natural\tilde\LLambda
+\delta^\sharp\tilde\LLambda \label{(4)}\fe}
in which $\delta^\sharp\tilde\LLambda$ denotes a part the would be 
inadmissible for the purpose of application of the variation principle, 
whereas $\delta^\natural\tilde\LLambda$ is a ``natural'' variation that 
would be allowed for this purpose. This means that for any (unperturbed) 
configuration that is ``on shell'' -- in the sense of satisfying the 
dynamical evolution equations provided by the constrained variation
 principle -- a generic admissible variation must satisfy the condition
{\be \delta^\natural\tilde\LLambda\cong 0\, ,\label{(5)}\fe}
using the symbol $\cong$ to indicate equivalence modulo the addition of 
a divergence (which, by Green's theorem, will give no contribution to the 
action integral from a variation that is compactly supported).

In the kind of medium with which we are concerned, the constrained 
variables will consist of the free and confined current vectors
$\nn_{\rmf}^{\,\mu}$ and  $\nn_{\rmc}^{\,\mu}$, while the unconstrained 
fields will consist of the triplet of scalar fields $\qq^{\rm _A}$ 
that are interpretable as local coordinates on the material base manifold, 
and from which, together with their gradient components 
$\qq^{\rm _A}_{\, ,\mu}$, an associated particle number (which might count 
ionic nuclei or crystalline solid lattice points in the relevant microscopic 
substructure) density current, $\nn_{\rm I}^{\,\nu}$ say, can be 
constructed in the manner prescribed below. The corresponding ``live'' part 
of the generic action variation will thus be expressible in the form 
{\be \delta^{_\heartsuit\!}\tilde\LLambda=\mmu^{\rmf}_{\ \mu}\delta 
\nn_{\rmf}^{\,\mu} +\mmu^{\rmc}_{\ \mu}\delta \nn_{\rmc}^{\,\mu}
+\LmP^\nu_{\,\rm_A}\delta \qq^{\rm _A}_{\, ,\nu}
+\frac{\partial \tilde\LLambda}{\partial \qq^{\rm _A}}\delta\qq^{\rm _A}
\, ,\label{(15)}\fe}
in which $\mmu^{\rmf}_{\ \mu}$ and  $\mmu^{\rmf}_{\ \mu}$ are interpretable 
as the 4-momentum covectors of the free and confined baryons respectively.

It is to be observed that the partial derivative $\LmP^\nu_{\,\rm_A}=
{\partial \tilde\LLambda}/{\partial \qq^{\rm _A}_{\, ,\nu}}$ has the
noteworthy property of being tensorial not just with respect to
transformations of the space time coordinates $\xx^\nu$ but also with 
respect to transformations $\qq^{\rm_A}\mapsto \tilde \qq{^{\rm_A}}$
%Changed rightarrow to \mapsto
of the material coordinates, $\qq^{\rm_A}$ whose effect will simply
be given by
{\be \qq^{\rm_A}\mapsto \tilde \qq{^{\rm_A}}\hskip 1 cm
\Rightarrow \hskip 1 cm\LmP^\nu_{\,\rm_A}\mapsto 
\LmP^\nu_{\,\rm_B}\,\frac{\partial \qq^{\rm_B}}{\partial \tilde
 \qq{^{\rm _A}}}\, ,\label{15a}\fe}
This good behaviour contrasts with the comportment of the remaining 
partial derivative, ${\partial\tilde\LLambda}/{\partial\qq^{\rm _A}}$,
for which the corresponding transformation has the non-tensorial form  
{\be\frac{\partial\tilde\LLambda}{\partial\qq^{\rm _A}}\mapsto\frac
{\partial\tilde\LLambda}{\partial\qq^{\rm _B}}\,\frac{\partial \qq^{\rm_B}}
{\partial \tilde \qq{^{\rm _A}}} +\LmP^\nu_{\,\rm_C}\,
\tilde\qq{^{\rm _B}_{,\nu}}\, \frac{\partial^2\qq^{\rm_C}}
{\partial \tilde \qq{^{\rm _B}}\partial \tilde \qq{^{\rm _A}}}
\, .\label{15b}\fe}

For each of the separate currents  $\nn_{\rmf}^{\,\mu}$ and  
$\nn_{\rmc}^{\,\mu}$ the admissible variations are generated
by flow world line displacements, which means that they will be 
specifiable by corresponding displacement vector fields, 
$\xi_{\rmf}^{\,\mu}$ and  $\xi_{\rmc}^{\,\mu}$ say,  as 
{\be \delta^\natural \nn_{\rmf}^{\,\mu}=-
%\vec\xi_{\rmf}\Libra
\Libra\lower 3.0pt \hbox{$\xi_{\rmf}$}
\, \nn_{\rmf}^{\,\mu}-\nn_{\rmf}^{\,\mu}\nabla_{\!\nu\,}
\xi_{\rmf}^{\,\nu}\, ,\hskip 1 cm
\delta^\natural \nn_{\rmc}^{\,\mu}=-
%\vec\xi_{\rmc}\Libra
\Libra\lower 3.0pt \hbox{$\xi_{\rmc}$}
\, \nn_{\rmc}^{\,\mu}-\nn_{\rmc}^{\,\mu}\nabla_{\!\nu\,}
\xi_{\rmc}^{\,\nu}\, ,\label{(6)}\fe}
in which the Lie derivative is just the commutator
{\be  
%\vec\xi_{\rmf}\Libra
\Libra\lower 3.0pt \hbox{$\xi_{\rmf}$}
\, \nn_{\rmf}^{\,\mu}=\xi_{\rmf}^{\,\nu}
\nabla_{\!\nu\,}\nn_{\rmf}^{\,\mu}-\nn_{\rmf}^{\,\nu}
\nabla_{\!\nu\,}  \xi_{\rmf}^{\,\mu}\, ,\hskip 1 cm
%\vec\xi_{\rmc}\Libra
\Libra\lower 3.0pt \hbox{$\xi_{\rmc}$}
\, \nn_{\rmc}^{\,\mu}=\xi_{\rmc}^{\,\nu}
\nabla_{\!\nu\,}\nn_{\rmc}^{\,\mu}-\nn_{\rmc}^{\,\nu}
\nabla_{\!\nu\,}  \xi_{\rmc}^{\,\mu}
\, .\label{(7)}\fe}
For the material base space, the corresponding the flow world line 
displacements will be given just by unrestricted variations of the base 
coordinates $\qq^{\rm_A}$, which will be expressible in the form
{\be \delta^\natural \qq^{\rm _A}=-\qq^{\rm _A}_{\, ,\mu}\xi^{\,\mu}
\, ,\label{(8)}\fe}
in terms of their own displacement vector field, $\xi^\mu$ say, which 
can be chosen independently a priori, but which will subsequently be 
subjected to the identification $\xi^{\,\mu}= \xi_{\rmc}^{\,\mu}$ when 
the confinement restriction (\ref{(2}) is imposed.

It can be seen (using integration by parts) that, in terms of the
displacement fields introduced by (\ref{(6)}) and (\ref{(8)}), the generic 
admissible variation for the current carrying medium will be expressible 
-- modulo a variationally irrelevant divergence term -- in the form
{\be  \delta^\natural\LLambda\cong -\ff^{\rmf}_{\, \nu}\,
\xi_{\rmf}^{\, \nu} - \ff^{\rmc}_{\, \nu}\,\xi_{\rmc}^{\, \nu}
-\ff^{\rms}_{\, \nu}\,\xi^{\nu}
\, ,\label{(9)}\fe}
in which the covectorial coefficient $\ff^{\rmf}_{\, \nu}$ will be 
interpretable as the force density acting on the free neutron current, 
and $\ff^{\rmc}_{\, \nu}$ will be interpretable as the force density 
acting on the confined baryon current, while the extra  coefficient 
$\ff^{\rms}_{\,\nu}$ is the supplementary force density due, as 
discussed below, to stratification or solid elasticity, that acts on the 
underlying ionic (crystalline or glass like) lattice structure. 

If we postulate that the system should obey the constrained variational 
principle that is expressible as the imposition of the ``on shell'' condition 
(\ref{(5)}), then it is evident from (\ref{(9)}) that the ensuing dynamical 
field equations will be expressible simply as the separate vanishing of each 
one of these force densities, i.e.  as the requirement that we should have 
$\ff^{\rmf}_{\, \nu}=0$, $\ff^{\rmc}_{\, \nu}=0$ and
$\ff^{\rms}_{\, \nu}=0$. We shall however be mainly concerned with other
possibilities, particularly cases in which one of the constituents, 
% with label {\srm X}=c 
% As abstract indecies for species are not used elsewhere I thought it was 
% unnessesary to introduce notation for it.
labelled ``c'' 
say, is  ``confined'' in the sense of being subject to
a convection condition of the form (\ref{(2})
which means that it has to move with the underlying ionic flow.
When the application of the variation principle is subject to
the corresponding convective constraint
{\be \xi_{\,\rmc}^{\, \nu}=\xi^{\nu}\, ,\label{11b}\fe}
the ensuing system of dynamical field equations will no longer entail
the separate vanishing of  $\ff^{\rms}_{\, \nu}$ and $\ff^{\rmc}_{\,\nu}$,
but only of  the amalgamated ionic force density $\ff^{\rmI}_{\, \nu}$ 
that is defined as their sum,
{\be \ff^{\rmI}_{\, \nu}= \ff^{\rms}_{\,\nu}+\ff^{\rmc}_\nu
\, ,\label{12}\fe}
which is interpretable as the net force density acting on the integrated 
system consisting of the ions in combination with the convected constituent.

\subsection{Noether identity for general covariance}

If we now replace the natural variations (\ref{(6)}) by the 
corresponding Lie variations
{\be \delta\nn_{\rmf}^{\,\nu}=-
%\vec\xi_\rmf\Libra 
\Libra\lower 3.0pt \hbox{$\xi_{\rmf}$}
\,\nn_{\rmf}^{\,\nu}
\, ,\hskip 1 cm
\delta\nn_{\rmc}^{\,\nu}=-
%\vec\xi_\rmc\Libra
\Libra\lower 3.0pt \hbox{$\xi_{\rmc}$} 
\,\nn_{\rmc}^{\,\nu}
\, ,\hskip 1 cm \delta\qq^{\rm _A}= -
%\vec\xi\Libra 
\Libra\lower 3.0pt \hbox{$\xi$}
\,\qq^{\rm _A} 
\, ,\label{(12)}\fe}
%removed \delta^\natural \qq^{\rm A} \, ,\label{(12)}\fe}
it can be seen that the current contributions will leave extra terms, 
so that we shall obtain a live variation of the form (\ref{(4)}) with 
$\delta^\natural\LLambda$ given by (\ref{(9)}) and with the extra 
``unnatural'' (variationally inadmissible) contribution given by
{\be \ \delta^\sharp\tilde\LLambda=\mmu^\rmf_{\,\nu}\delta^\sharp
\nn_{\rmf}^{\,\nu}+\mmu^\rmc_{\,\nu}\delta^\sharp\nn_{\rmc}^{\,\nu}
\, ,\hskip 1 cm \delta^\sharp\nn_{\rmf}^{\,\nu}=\nn_{\rmf}^{\,\nu}
\nabla_{\!\mu}\xi_\rmf^{\,\mu}\, ,\hskip 1 cm\delta^\sharp
\nn_{\rmc}^{\,\nu}=\nn_{\rmc}^{\,\nu}\nabla_{\!\mu}\xi_\rmc^{\,\mu}
\, .\label{(13)}\fe}
We can now obtain a Trautman type Noether identity (that will hold 
regardless of whether or not the variational field equations are 
satisfied) by taking the variations of all the relevant fields, ``live'' 
as well as ``dead'', to be given by the action of the same displacement 
field $\xi^\nu$ whose effect on the background was given by (\ref{(3)}),
so that the resulting effect on the Lagrangian scalar will be given 
just by the corresponding displacement variation
{\be \delta\tilde\LLambda=-\xi^\nu\nabla_{\!\nu}\tilde\LLambda
\cong\tilde\LLambda\nabla_{\!\nu}\xi^\nu\, .\label{(10)}\fe}
We thereby obtain a relation of the form
{\be  \xi^\nu\left(\ff^{\rmf}_{\,\nu}
+\ff^{\rmc}_{\, \nu} +\ff^{\rms}_{\, \nu}\right) \cong
\delta^\ddagger\tilde\LLambda+\delta^\sharp\tilde\LLambda-
\tilde\LLambda\nabla_{\!\nu}\xi^\nu\, .\label{(14)}\fe}
in which,  by  (\ref{(1)}), (\ref{(3)}) and (\ref{(13)}), the ``dead''
and ``unnatural'' parts will be respectively given by
{\be  \delta^\ddagger\tilde\LLambda=-2\frac{\partial\tilde\LLambda}{\partial
\gr_{\mu\nu}}\nabla_{\!\mu}\xi_\nu\, ,\hskip 1 cm\delta^\sharp\tilde\LLambda
=(\mmu^\rmf_{\,\nu}\nn_{\rmf}^{\,\nu}+\mmu^\rmf_{\,\nu}
\nn_{\rmf}^{\,\nu})\nabla_{\!\mu}\xi^\mu\, .\label{(11)}\fe}
The relation (\ref{(14)}) can thereby be rewritten in the standard form
{\be \xi^\nu\left(\ff^{\rmf}_{\,\nu}+ \ff^{\rmc}_{\,\nu}+
\ff^{\rms}_{\,\nu}\right)\cong -\TT^\mu_{\ \nu}\nabla_{\!\mu}
\xi^\nu\cong\xi^\nu\nabla_{\!\mu}\TT^\mu_{\ \nu}\, ,\label{(16)}\fe}
in which the  geometric stress momentum energy density tensor can be read 
out as
{\be \TT^{\mu\nu}=2\frac{\partial\tilde\LLambda}
{\partial\gr_{\mu\nu}}  +\PPsi\gr^{\mu\nu}\, ,\hskip 1 cm
\PPsi=\tilde\LLambda-\mmu^{\rmf}_{\ \nu}\nn_{\rmf}^{\,\nu}
-\mmu^{\rmc}_{\ \nu} \nn_{\rmc}^{\,\nu}\, . \label{(18)}\fe}
Since the arbitrary field $\xi^\nu$ can be taken to be non zero only
in the immediate neighbourhood of any chosen point, the Noether identity
(\ref{(16)}) implies that, at each point, this stress energy tensor
will satisfy a divergence identity of the simple form
{\be \nabla_{\!\mu}\TT^\mu_{\ \nu}=\ff^{\rmf}_{\,\nu}+\ff^{\rmc}_{\,\nu}
+\ff^{\rms}_{\,\nu}\, .\label{(19)}\fe}
In the ``on shell'' case for which the variational evolution equations are
satisfied, the force density contributions on the right
of (\ref{(19)}) will all drop out. 

\subsection{Canonical formulation}

%Sentence changed below:
If, instead of working out the variations in terms of force densities modulo a
divergence as in (\ref{(14)}) and (\ref{(16)}), we insert the displacement 
contributions (\ref{(11)}) and (\ref{(12)}) into the ``live'' variation 
(\ref{(15)}) directly in (\ref{(0)}) we get an identity in which the 
coefficients of the locally adjustable fields $\xi^\mu$ and 
$\nabla_{\!\mu}\xi^\nu$ must vanish separately. The former of these 
conditions reduces to a triviality, but the latter provides a relation
showing that the geometrically defined stress energy tensor (\ref{(19)}) 
can be rewritten in the equivalent canonical form
{\be \TT^\mu_{\ \nu}=\PPsi\ddelta^\mu_{\ \nu}+\mmu^{\rmf}_{\  \nu}
\nn_{\rmf}^{\,\mu}+\mmu^{\rmc}_{\  \nu}\nn_{\rmc}^{\,\mu}
-\LP^\mu_{\ \nu}\, ,\label{(21)}\fe}
in which the stress contribution at the end is given in terms
of the coefficient introduced just before (\ref{15a}) by
{\be \LP^\mu_{\ \nu}=\LmP^\mu_{\,\rm _A}\qq^{\rm _A}_{\, ,\nu}
\, .\label{21a}\fe}

The ``live'' variation formula (\ref{(15)}) also provides what is needed 
for the derivation of the corresponding expressions for the force densities
introduced in (\ref{(9)}) which for the fluid constituents will have the 
form that is already familiar from the preceding work~\cite{C89}. In 
terms of the corresponding generalised vorticity 2-forms defined, using 
square brackets for index antisymmetrisation, by
{\be \ww^{\rmf}_{\,\mu\nu}=2\nabla_{\![\mu}\mmu^{\rmf}\!{_{\!\nu]}}
\, ,\hskip 1 cm\ww^{\rmc}_{\,\mu\nu}=2\nabla_{\![\mu}
\mmu^{\rmc}\!{_{\!\nu]}}\, ,\label{(22a)}\fe}
these current force densities will be given by
{\be \ff^{\rmf}_{\, \nu}= \nn_{\rmf}^{\,\mu} 
\ww^{\rmf}_{\,\mu\nu}+\mmu^{\rmf}_{\ \nu} \nabla_{\!\mu}
\nn_{\rmf}^{\,\mu}\, ,\hskip 1 cm \ff^{\rmc}_{\, \nu}= 
\nn_{\rmc}^{\,\mu} \ww^{\rmc}_{\,\mu\nu}+\mmu^{\rmc}_{\ \nu} 
\nabla_{\!\mu}\nn_{\rmc}^{\,\mu}\, ,\label{(22)}\fe}
while the force density acting on the underlying atomic 
structure of the medium will be given by the formula
{\be \ff^{\rms}_{\,\mu}=\frac{\delta \tilde\LLambda}{ \delta \qq^{\rm _A}}
\qq^{\rm _A}_{\, ,\mu}\, ,\label{(23)}\fe}
in which the Eulerian derivative is given by a prescription of the
usual form
{\be \frac{\delta \tilde\LLambda}{ \delta \qq^{\rm _A}}=\frac{\partial\tilde\LLambda}
{\partial  \qq^{\rm _A}}-\nabla_{\!\nu}\LmP^\nu_{\,\rm _A}
\, ,\hskip 1 cm \LmP^\nu_{\,\rm _A}=\frac{\partial\tilde\LLambda}
{\partial  \qq^{\rm _A}_{\, ,\nu}} \, .\label{(24)}\fe}
It is important to notice that the non-tensorial base coordinate
transformation property  (\ref{15b}) of the first term in this formula 
will be cancelled by that of the second, so that the effect on the Eulerian
derivative of a change of the material base coordinates will be
given simply by
{\be \qq^{\rm_A}\mapsto \tilde \qq{^{\rm_A}}\hskip 1 cm \Rightarrow 
\hskip 1 cm \frac{\delta\tilde\LLambda}{\delta \qq^{\rm_A}}
\mapsto  \frac{\delta\tilde\LLambda}{\delta \qq^{\rm_B}}
\,\frac{\partial \qq^{\rm_B}}{\partial \tilde \qq{^{\rm _A}}}
\label{24a}\, ,\fe}
which shows that the (stratification and solid elasticity) force density 
(\ref{(23)}) is invariant with respect to such transformations, an 
important property that is not so obvious from its detailed expression
{\be \ff^{\rms}_{\,\mu}=\frac{\partial \tilde\LLambda}{ \partial \qq^{\rm _A}}
\qq^{\rm _A}_{\, ,\mu}+ \LmP^\nu_{\,\rm _A} \nabla_{\!\mu}
\qq^{\rm _A}_{\, ,\nu}-\nabla_{\!\nu}\left( \LmP^\nu_{\,\rm _A}
\qq^{\rm _A}_{\, ,\mu}\right)\, ,\label{(25)}\fe}
in which only the final term is separately invariant.

It is important to observe that, regardless of what -- variational or
other -- dynamical field equations may be imposed, the force density 
(\ref{(25)}) will never be able to do any work on the medium since, as an 
obvious consequence of (\ref{(23)}) it must automatically satisfy the 
further identity 
{\be \ff^{\rms}_{\,\nu} \uu^\nu=0\, ,\label{(26)}\fe}
where $\uu^\mu$ is the unit 4-velocity of the medium as specified by the 
defining conditions
{\be \qq^{\rm _A}_{\, ,\nu} \uu^\nu=0\, ,\hskip 1 cm \uu_\nu \uu^\nu=
-\cc^2\, ,\label{(27)}\fe}
where $\cc$ is an adjustable parameter that is introduced to
facilitate the derivation \cite{CCC} of the Newtonian limit case.

If the variation principle (\ref{(5)}) were imposed for the three 
independent displacement fields, so that -- as remarked above -- by 
(\ref{(9)}) the three separate force densities $\ff^{\rmf}_{\,\mu}$, 
$\ff^{\rmc}_{\,\mu}$ and $\ff^{\rms}_{\, \mu}$ would all have to 
vanish, then it evidently follows from the Noether identity 
(\ref{(19)}) that the divergence condition 
{\be \nabla_{\!\mu}\TT^\mu_{\ \nu}=0
\, ,\label{(28)}\fe}
would have to hold. This condition is interpretable as an 
 energy-momentum balance equation that must be satisfied whenever the 
system is not subject to any external forces. So long as we are
concerned only with a system that is isolated in this sense, so that the 
energy-momentum balance condition (\ref{(28)}) is satisfied, it follows 
from (\ref{(19)}) that, even for a  more general model admitting the 
presence of dissipative or other internal forces, these must be be 
specified in such a way as to satisfy the total force balance condition
{\be \ff^{\rmf}_{\, \nu}+ \ff^{\rmc}_{\,\nu}+ \ff^{\rms}_{\,\nu}
=0\, ,\label{(29)}\fe} 
in order for the model to be self consistent. More particularly,
in view of (\ref{(26)}) it can be seen that the fluid force densities
by themselves will have to satisfy the condition
{\be \uu^\nu(\ff^{\rmf}_{\,\nu}+ \ff^{\rmc}_{\,\nu})
=0\, ,\label{(30)}\fe} 
expressing energy balance in the local rest frame of the underlying
medium.

\section{Conservative dynamical conditions}

\subsection{Baryon conservation and the confinement constraint}

The system under consideration is designed for applications in which 
the total baryon current
{\be  \nn_{\rm b}^{\,\nu}= \nn_{\rm c}^{\,\nu} +  \nn_{\rm f}^{\,\nu}
\, ,\label{(31)}\fe}
can be decomposed as the combination of a free neutron current
$\nn_{\rm f}^{\,\nu}$ and a convected part $\nn_{\rm c}^{\,\nu}$
consisting of the remaining baryons (some of the neutrons and all
of the protons) that is convected with the nuclei, so that it is
characterised by the confinement condition
{\be \nn_\rmc^{\,\nu} =\nn_\rmc \uu^\nu\, , \hskip 1 cm
\nn_\rmc=(-\nn_{\rmc\nu}\nn_\rmc^{\,\nu})^{1/2}/\cc\, .\label{33a} \fe}

As a consequence of the condition of conservation of the total
baryon current, 
{\be \nabla_{\!\nu}\nn_{\rm b}^{\,\nu}=0 \, , \label{(32)}\fe}
it follows that in terms of the ``free'' current vorticity 2-form 
$\ww^{\rmf}_{\,\mu\nu}$ the energy conservation identity (\ref{(30)}) 
will reduce to the form
{\be \uu^\mu \ww^{\rm f}_{\,\mu\nu}\nn_{\rm f}^{\,\nu}=
\uu^\mu\left(\mmu^{\rm f}_{\ \mu}-\mmu^{\rm c}_{\ \mu}\right)\nabla_{\!\nu}
\nn_{\rm f}^{\, \nu}\, .\label{(33)}\fe}

Due to the confinement constraint, there is a freedom of adjustment 
in the specification of the confined particle momentum covector
$ \mmu^{\rm c}_{\ \mu}$  by the general formula (\ref{(15)}), but this 
does not affect the quantities involved in (\ref{(33)}) because there is 
no ambiguity in the specification of the energy component 
{\be {\calE}^{\rm c}=- \uu^\mu\,\mmu^{\rm c}_{\ \mu} \, .\label{(34)}\fe}
An obvious way of resolving the ambiguity in the remaining (space) 
components of the confined particle momentum  would be to set them to zero, 
so that the complete covector would simply take the form 
$\mmu^{\rm c}_{\ \mu}=-{\calE}^{\rm c} \, \uu_{\mu}$. Such a choice would be 
adequate for the specialised purpose matching the formalism used for the 
non-conducting solid limit case \cite{CCC}. However for the broader 
objective of consistency with the formalism developed for the 
multiconstituent fluid limit case \cite{CK92} we shall adopt a 
more satisfactory alternative ansatz for resolving the ambiguity in the 
specification of $ \mmu^{\rm c}_{\ \mu}$, which is to fix it by imposition 
of a convention to the effect that
{\be  \LmP^\nu_{\,\rm_A}\, \uu_\nu=0\, .\label{(34a)}\fe}

\subsection{Vortex pinning and chemical equilibrium}
\label{pineq}

As the condition of superfluidity of the ``free'' constituent will 
consist in a restriction of the possible behaviour of the corresponding
momentum covector  $\mmu^{\rmf}_{\ \mu}$, it is important to remark that
this covector will not be affected by the chemical basis changes that
will be discussed in Subsection \ref{chemadj}. On a mesoscopic scale 
(large compared with the microscopic lattice spacing but small compared 
with intervortex separation) the condition of superfluidity will be 
expressed by the condition that this covector should have the form of a 
gradient, $\mmu^{\rmf}_{\ \mu}=\hhbar\nabla_{\!\mu}\varpphi$ of a periodic 
quantum phase angle $\varpphi$, with the corollary that the corresponding 
``free'' vorticity tensor $\ww^{\rmf}_{\,\mu\nu}$ will also vanish at the 
mesoscopic level. At the macroscopic level, on a scale large compared with 
the separation between vortices, this vorticity will have a large scale 
average value that will not vanish, but to be compatible with a mesoscopic 
fibration by 2 dimensional vortex lines it must still be algebraically 
restricted by the degeneracy condition
{\be \ww^{\rm f}_{\,\mu[\nu}\ww^{\rm f}_{\,\rho\sigma]}=0
\, .\label{(36)}\fe}

This condition will automatically be satisfied by the field equations that 
will result from the relevant variation principle, which will require that 
the vanishing perturbation condition (\ref{(5)}) should be satisfied, not 
for independent variations of all three of the displacement fields 
$\xi_{\rm f}^{\,\mu}$,  $\xi_{\rm c}^{\,\mu}$, $\xi^\mu$, which in view of 
the constraint (\ref{(31)}) would lead to overdetermination, but just for 
variations satisfying the corresponding restraint (\ref{11b}). This 
requirement entails the vanishing, not of all three of the relevant force 
densities $\ff^{\rmf}_{\,\mu}$, $\ff^{\rmc}_{\,\mu}$, $\ff^{\rms}_{\,\mu}$ 
given by the prescriptions (\ref{(22)}) and (\ref{(25)}), but only of the 
free force density $\ff^{\rmf}_{\,\mu}$  and of the amalgamated force 
density $ \ff^{\rmI}_{\,\mu}$ given by (\ref{12}) that acts on the part that 
is convected with the ionic lattice. 

According to the general formula (\ref{(22)}), the vanishing of the free 
force density $\ff^\rmf_{\,\mu}$ entails not only the separate conservation 
of the free part $\nn_\rmf^{\,\mu}$ of the current density but also a 
dynamic equation of the familiar form
{\be \ww^{\rm f}_{\,\mu\nu}\nn_{\rm f}^{\,\nu}=0\, ,\label{(38)}\fe}
which is interpretable as meaning that the vortex lines are ``frozen'' into
the free fluid in the sense of being dragged along with it. 

Instead of postulating the strict application of the variation principle,
we can obtain a non-dissipative model of another physically useful kind
by postulating that the vortices are ``pinned'' in the sense of being frozen, 
not into the fluid but into the atomic lattice structure, which means 
replacing (\ref{(38)}) by a dynamical equation of the analogous form 
 {\be \ww^{\rm f}_{\,\mu\nu}u^{\nu}=0\, ,\label{(39)}\fe}
which unlike (\ref{(38)}) is unaffected by changes of the chemical
basis \cite{CK92} that is used, as discussed in Subsection \ref{chemadj},
to specify which of the neutrons are to be considered to be ``free''.
 
Whichever of the alternative possibilities (\ref{(38)}) or (\ref{(39)}) is 
adopted, the degeneracy condition (\ref{(36)}) will evidently be satisfied,
 and more particularly the left hand side of the identity (\ref{(33)}) will 
vanish, which means that the model must be such as to satisfy the condition
{\be \uu^\mu\left(\mmu^{\rm f}_{\ \mu}-\mmu^{\rm c}_{\ \mu}\right)
\nabla_{\!\nu}\nn_{\rm f}^{\, \nu}=0\, .\label{(40)}\fe}
In the strictly variational case this requirement  will obviously  be 
implemented by the separate conservation condition
{\be \nabla_{\!\nu}\nn_{\rm f}^{\, \nu}=0\, ,\label{(41)}\fe}
which will be physically realistic in the context of high frequency 
oscillations. However in applications to slow long term variations it will be 
more realistic to use a model based on the alternative possibility, namely 
that of chemical equilibrium, as given by the relation
{\be  \uu^\mu\left(\mmu^{\rm f}_{\ \mu}-\mmu^{\rm c}_{\ \mu}\right)=0\, ,
\label{(42)}\fe}
which, like (\ref{(39)}) is  does  not depend on which fraction
of the neutrons is considered to be  ``free''.

Thus, depending on how we choose between the alternatives (\ref{(38)}) or 
(\ref{(39)}), and between the alternatives (\ref{(41)}) or (\ref{(42)}), we 
can use the same Lagrangian master function $\tilde\LLambda$ for the 
specification of 4 different kinds of non-dissipative models, which are 
categorisable as unpinned with separate conservation or chemical equilibrium, 
and pinned with separate conservation or chemical equilibrium.

\section{Internal action function}

\subsection{Subtraction of the ballistic contribution}

For the purpose of studying the Newtonian limit it is convenient
\cite{CCC} to split up the total action function into parts depending
just on  the confined number density $\nn_\rmc$ defined by
(\ref{33a}) and the corresponding free number density $\nn_\rmf$
defined by setting
{\be \nn_\rmf^{\,\nu} =\nn_\rmf \uu_\rmf^{\,\nu}\, , \hskip 1 cm
\nn_\rmf=(-\nn_{\rmf\nu}\nn_\rmf^{\,\nu})^{1/2}/\cc\, ,\label{33b} \fe}
in a decomposition of the form
{\be \tilde\LLambda=\tilde\LLambda_{\rm_{bal}}+
\LLambda_{\rm_{int}}\, ,\label{49}\fe}
(using the tilde just to distinguish the parts that will be relatively 
large in the non relativistic limit) in which the first term is a ballistic 
contribution (meaning that by itself it would simply imply geodesic motion 
of the separate contributions) given in terms of a fixed mass parameter 
$\mm$ by an expression of the simple form
{\be \tilde\LLambda_{\rm_{bal}}=-\mm\,\cc^2(\nn_\rmc + \nn_\rmf) 
\, ,\label{50}\fe}
so that all the non-trivial information about the system will be contained 
in the remaining internal action contribution $\LLambda_{\rm_{int}}$, 
which remains finite in the large $\cc$ limit \cite {CCC} (for which the 
tilde parts would diverge).  
 
As far  as the relativistic treatment is concerned, $\mm$ may be given 
an arbitrarily chosen positive value or simply set to zero. The reason 
for taking the trouble of introducing it at all is that in appropriate 
circumstances $\mm$ can be judiciously chosen in such a way as to make 
the internal remainder  $\LLambda_{\rm_{int}}$ relatively small, so that 
it can be dealt with as a small perturbation in an approximation of 
Newtonian type, for which $\mm$ will be interpreted as the ``rest mass 
per baryon''. In most astrophysical contexts for which a Newtonian 
approximation is applicable, and in particular for local perturbations 
in the crust of a neutron star, it will be good enough to use the 
proton mass  calibration $\mm=\mmm_{\rm p}$, but it is well known
that under laboratory conditions for which higher accuracy is needed 
it will be better to take the ``rest mass per baryon'' to be the 
standard atomic unit defined as a sixteenth of the mass of an ordinary 
$O^{16}$ oxygen atom. At the opposite extreme, at extremely high
densities it might seem more natural to interpret the ``rest mass per 
baryon'' as meaning three times the ``bare mass'' of a quark, but 
subtraction of the relatively small rest mass contribution defined in 
that way would be pointless. In practise, under the highly relativistic 
conditions prevailing in the core of a neutron star, the contribution 
from $\LLambda_{\rm_{int}}$ will inevitably be relatively large, so 
that the most judicious choice may simply be to set $\mm=0$.

However $\mm$ is chosen, its introduction will lead to a corresponding 
momentum decomposition of the form
{\be \mmu^\rmc_{\,\nu}=\mm \,\uu_\nu+\cchi^\rmc_{\,\nu}\hskip
1 cm\mmu^\rmf_{\,\nu}=\mm\, \uu_{\rmf\, \nu}+\cchi^\rmf_{\,\nu}
\, ,\label{51}\fe}
in which the non-trivial ``chemical'' momentum contributions from the
internal part of the action can be read out from its expansion
in the form
{\be \delta^{_\heartsuit\!}\LLambda_{\rm_{int}}=\cchi^{\rmf}_{\ \mu}\delta 
\nn_{\rmf}^{\,\mu} +\cchi^{\rmc}_{\ \mu}\delta \nn_{\rmc}^{\,\mu}
+\LmP^\nu_{\,\rm_A}\delta \qq^{\rm _A}_{\, ,\nu}
+\frac{\partial \LLambda_{\rm_{int}}}{\partial \qq^{\rm _A}}
\delta\qq^{\rm _A} \, ,\label{52}\fe}
in which (since the specification of the ballistic part does not
involve the $\qq^{\rm A}$) it can be checked that the coefficients in 
the last two terms will be the same as those already introduced in the 
analogous expansion (\ref{(15)}) of the total action density. 
 
\subsection{Material projections}

According to the general principle developed in the preceding
work \cite{C89} the internal action will be expressible, at a 
given material position, as characterised by the material
coordinates $\qq^{\rm_A}$, as a function just of the (time
dependent) induced metric with contravariant components
given by
{\be\gamm^{\rm_{AB}}=\gr^{\mu\nu}\qq^{\rm_A}_{,\mu}\qq^{\rm_B}_{,\nu}
=\gamm^{\mu\nu}\qq^{\rm_A}_{,\mu}\qq^{\rm_B}_{,\nu}\, ,\hskip 1 cm
\gamm^{\mu\nu}=\gr^{\mu\nu}+\cc^{-2}\uu^\mu\uu^\nu\, ,\label{53}\fe}
and of the relative current components
{\be \nn^{\rm_A}=\nn_\rmf^{\,\nu}\qq^{\rm_A}_{,\nu}=\nn_{^\perp}^{\,\nu}
\qq^{\rm_A}_{,\nu} \, ,\hskip 1 cm  \nn_{^\perp}^{\,\nu}=
\gamm^\nu_{\, \mu}\nn_\rmf^{\,\mu}= \nn_{^\Vert}[\vv]^\nu
\, ,\hskip 1 cm [\vv]^\nu \uu_\nu=0
\, ,\label{53a}\fe}
together with the corresponding material rest frame number densities,
namely the confined density $\nn_\rmc$ for which no projection is
needed, and the  free number density given in terms of the Lorentz 
factor $\gamm$ specified by the relative current velocity $[\vv]^\mu$ as
{\be \nn_{^\Vert}=-\cc^{-2}\uu_\nu\nn_\rmf^{\,\nu}=\gamm\,\nn_\rmf
\, ,\hskip 1 cm \gamm=(1-[\vv]^2/\cc^2)^{-1/2}\, .\label{53b}\fe}

These conditions imply that the internal action density (which will be 
the same as the complete action density if we choose $\mm=0$) will have
a generic variation of the form
{\be \delta\LLambda_{\rm_{int}}=-\cchi^\rmc\,\delta  \nn_{\rm c}
-\cchi^{_\Vert}\,\delta  \nn_{^\Vert}+\pp^{_\perp}_{\, \rm_A}\delta  
\nn^{\rm_A}+\frac{\partial\LLambda_{\rm_{int}}}{\partial 
\gamm^{\rm_{AB}}}\delta  \gamm^{\rm_{AB}}+\frac{\partial
\LLambda_{\rm_{int}}}{\partial \qq^{\rm_A}}\delta \qq^{\rm_A}
\, .\label{58}\fe}
This provides an associated ``convective'' variation \cite{C89}
(in which, with respect to appropriately dragged coordinates,
both $\qq^{\rm_A}$ and $\qq^{\rm_A}_{,\nu}$ 
%removed delta
are held 
constant) of the form
{\be \delta_{\!\rm _c}\LLambda_{\rm_{int}}=-\cchi^\rmc\,\delta_{\!\rm _c} 
\nn_{\rmc}-\cchi^{_\Vert}\,\delta_{\!\rm _c}  \nn_{^\Vert}+
\frac{\partial\LLambda_{\rm_{int}}}{\partial \nn_{^\perp}^{\,\nu}}
\delta_{\!\rm _c}  \nn_{^\perp}^{\,\nu}+\frac
{\partial\LLambda_{\rm_{int}}}{\partial \ggamma^{\mu\nu}}
\delta_{\!\rm _c}  \ggamma^{\mu\nu}\, ,\label{80}\fe}
in terms of tensorial coefficients defined by 
{\be \frac{\partial\LLambda_{\rm_{int}}}{\partial \nn_{^\perp}^{\,\nu}}
=\pp^{_\perp}_{\ \nu}=\pp^{_\perp}_{\, \rm_A}\qq^{\rm_A}_{,\nu}
\, ,\hskip 1 cm\frac{\partial\LLambda_{\rm_{int}}}
{\partial\ggamma^{\mu\nu}}=\frac{\partial\LLambda_{\rm_{int}}}
{\partial \gamm^{\rm_{AB}}}\qq^{\rm_A}_{,\nu}\qq^{\rm_B}_{,\mu}
\, .\label{81}\fe}

\subsection{Ionic density and chemical gauge adjustments}
\label{chemadj}

It is computationally convenient and physically natural to take the material
base space to be endowed with a measure form that is specifiable in
terms of  antisymmetric components, $\nn_{\rm I_{ABC}}$ say, that are 
fixed in the sense of depending only on the $\qq^{\rm_A}$, and that will 
determine a corresponding scalar space-time field $\nn_{\rm I}$ by the 
determinant formula
{\be \nn_{\rm I}^{\,2}=\frac{1}{3!} \nn_{\rm I_{ABC}} \nn_{\rm I_{DEF}}\,
\gamm^{_{AD}}\gamm^{_{BE}}\gamm^{_{CF}} \, .\label{55}\fe}
For purposes of physical interpretation it will usually be convenient to 
take this measure to represent the number density of ionic nuclei, so that 
the confined baryon number density $\nn_\rmc=-c^{-2}\uu_\nu\nn_\rmc^{\, \nu}$
% $\nn_\rmc=\tt_\nu\nn_\rmc^{\, \nu}$
% $\tt_nu$ not defined or used elsewhere 
will be given by
{\be \nn_{\rm c}={\Armc}\nn_{\rm I}\, ,\label{56}\fe}
where ${A_\rmc}$ is the atomic number, meaning the number of confined 
baryons (protons plus confined neutrons) per nucleus. It is to be observed 
that, whereas the number current $\nn_{\rm c}^{\,\nu}$ will fail to be 
conserved in the chemical equilibrium case characterised by (\ref{(42)}), 
the formalism is such that the corresponding ionic number current
{\be \nn_{\rm I}^{\,\nu}=\nn_{\rm I}\, \uu^\mu\fe}
must  automatically satisfy the conservation law
{\be \nabla_{\!\nu} \nn_{\rm I}^{\,\nu}=0\label{57}\fe}
as a mathematical identity. Although there can be exceptions (for example 
when nuclear fission occurs) it will in fact most commonly be realistic 
to suppose that the number of ionic nuclei is indeed conserved even when 
the number of protons and neutrons they contain is changing (whether by 
transfusive interchange of the confined part or just by conservative 
``dripping'' of neutrons into the ambient superfluid).

For particular physical applications, the specification of the part of the
baryon current that is to be treated as ``confined'' will be to some extent
dependent on the timescales of the processes under consideration as compared
with with the timescales for quantum tunnelling through the confining 
barriers containing the nuclei. For relatively rapid processes the fraction
of the relevant quantum states that should be considered to be confined may 
be subject to an increase, so the corresponding current, $\hat\nc^\nu$
say, of baryons that are effectively confined will be somewhat larger than
the value, $\nc^\nu$, that would be appropriate for processes occurring
over longer timescales. It is therefore of interest to consider the effect 
of such a chemical basis adjustment, as given by a transformation of the
form
{\be \hat\nc{^\nu}=(1+\epsilon)\nc^\nu\, ,\hskip 1 cm
\hat\nf{^\nu}=\nf^\nu-\epsilon\nc^\nu\, ,\label{44}\fe}
for some suitably specified dimensionless adjustment parameter $\epsilon$. 
So long as $\epsilon$ is just a constant, it is immediately apparent,
that such a transformation will leave the
``free'' 4-momentum covector invariant, meaning that we shall have
{\be \mmu^\rmf_{\,\nu}=\hat\mmu{^\rmf_{\,\nu}} \, ,\label{45}\fe}
and,  as in the fluid case \cite{CK92} it is also easy to see that 
the entire  stress energy tensor(\ref{(21)}) will be similarly invariant,
{\be \TT^\mu_{\ \nu}=\hat\TT{^\mu_{\ \nu}}\, .\label{46}\fe}

A less obvious observation \cite{CCH05} is that this covariance property of 
the canonical stress energy tensor is not restricted to transformations for
which $\epsilon$ is constant. It will still hold whenever $\epsilon$ is 
specified as a function just of the confined atomic number
$\Armc=\nn_\rmc/\nn_\rmI$, and of the material position coordinates 
$\qq^{\rm_A}$. Since it can be seen from (\ref{55}) that the variation of
the ionic number will have a ``live'' (fixed background) part expressible
in the form
{\be\delta^{_\heartsuit\!}\nn_{\rmI}=\nn_{\rm I}\, \gamm_{\rm_{AB}}
\qq^{\rm _B}_{\, ,\mu} \gamma^{\mu\nu}\delta\qq^{\rm _A}_{\, ,\nu}
+\frac{\partial\nn_\rmI}{\partial  \qq^{\rm _A}}\delta\qq^{\rm_A}
\, ,\label{62}\fe}
the corresponding variation of the adjustment parameter will have the form
{\be\delta^{_\heartsuit\!}\epsilon=\frac{\partial\epsilon}
{\partial \Armc}\left(\frac{\delta^{_\heartsuit\!} \nc}{\nn_\rmI}
-\Armc\gamm_{\rm_{AB}} \qq^{\rm _B}_{\, ,\mu}
\gamma^{\mu\nu}\delta\qq^{\rm _A}_{\, ,\nu} \right)
+\left(\frac{\partial\epsilon}{\partial  \qq^{\rm _A}}
-\frac{\partial \epsilon}{\partial \Armc}\frac{\Armc}{\nn_\rmI}
\frac{\partial \nn_\rmI}{\partial \qq^{\rm _A}}
\right)\delta\qq^{\rm_A}\, .\label{47}\fe}
Unlike the free particle momentum which is subject to the chemical 
invariance condition (\ref{45}), the confined particle momentum 
covector will undergo a non-trivial chemical adjustment given by
{\be \mmu^\rmc_{\,\nu}=\hat\mmu{^\rmc_{\,\nu}}+(\hat\mmu{^\rmc_{\,\mu}}
-\hat\mmu{^\rmf_{\,\mu}})\left(\epsilon\,\delta^\mu_\nu-
\frac{\Armc}{\cc^2}\frac{\partial\epsilon}{\partial \Armc}\uu^\mu
\uu_\nu\right)\, .\fe}
However it can be verified that the extra terms will cancel out in the 
formula for the stress energy tensor so that the invariance condition 
(\ref{46}) will remain valid.

\subsection{Evaluation of momenta and stress energy tensor}

To proceed from (\ref{58}), what we need is the corresponding ``live'' 
variation, as carried out at a fixed position in a fixed background, 
for which one  can obtain an expression of the standard form
{\be \delta^{_\heartsuit\!}\LLambda_{\rm_{int}}=
\cchi^\rmc_{\,\nu}\,\delta\nc^\nu+\cchi^\rmf_{\,\nu}\,\delta\nf^\nu
+\LmP^{\, \nu}_{\,\rm_{A}}\,\delta\qq^{\rm_A}_{,\nu}+\frac{\partial
\LLambda_{\rm_{int}}}{\partial \qq^{\rm_A}}\,\delta \qq^{\rm_A} 
\, ,\label{83}\fe}
in which, by using  the constraint $\uu^\nu\delta\qq^{\rm_A}_{,\nu}
=-\qq^{\rm_A}_{,\nu}\delta\uu^\nu$ the terms can be recombined in such a 
way as to obtain a well defined set of coefficients characterised by the
requirement that $\LmP^{\, \nu}_{\,\rm_{A}}$ should satisfy the 
condition (\ref{(34a)}). It can be seen that the required (purely 
spacelike) value of  $\LmP^{\, \nu}_{\,\rm_{A}}$ will be given by
{\be\LmP^{\, \nu}_{\,\rm_{A}}=\pp^{_\perp}_{\,\rm_A}
\,\nn_{^\perp}^{\,\nu}+2\,\frac{\partial
\LLambda_{\rm_{int}}}{\partial \gamm^{\rm_{AB}}}\,\qq^{\rm_B}_{,\mu}
\ggamma^{\mu\nu} \, ,\label{84}\fe}
while the corresponding expressions for the internal contributions to 
the free and confined and particle 4-momentum covectors will be given by
{\be\cchi^\rmf_{\,\nu}=\frac{\cchi^{_\Vert}}{\cc^2} \, \uu_\nu
+ \pp^{_\perp}_{\ \nu}\, ,\hskip 1 cm \cchi^\rmc_{\,\nu}
=\frac{\cchi^\rmc}{\cc^2} \, \uu_\nu-\frac{\nn_{^\Vert}}{\nc} \,
\pp^{_\perp}_{\ \nu}+\frac{\cchi^{_\Vert}}{\cc^2\nn_\rmc}
\, \nn_{^\perp\nu}\, , \label{85}\fe}
which implies that we shall have
{\be \cchi^{_\Vert}=-\uu^\nu\cchi^\rmf_{\,\nu}\, ,\hskip 1 cm
\cchi^\rmc=-\uu^\nu\cchi^\rmc_{\,\nu}\, .\label{86}\fe}

It follows, according to the canonical formula (\ref{(21)}), that the 
internal contribution to the stress energy tensor will be given by
{\be \TT^{\,\mu}_{\!_{\rm int}\nu}=\cchi^{\rmf}_{\  \nu}\nf^{\,\mu}
+\cchi^{\rmc}_{\  \nu}\nc^{\,\mu}+\PPsi\ddelta^\mu_{\ \nu}
-\LP^\mu_{\ \nu}\, ,\label{87}\fe}
with
{\be\LP^{\mu}_{\ \nu}=\pp^{_\perp}_{\ \nu}
\,\nn_{^\perp}^{\,\mu}+2\,\frac{\partial
\LLambda_{\rm_{int}}}{\partial \ggamma^{\nu\rho}}\,\ggamma^{\rho\mu} 
\, ,\label{87a}\fe}
and with the generalised pressure scalar $\PPsi$ given, as in the 
multiconstituent fluid case, by
{\be \PPsi=\LLambda_{\rm_{int}}-\cchi^{\rmf}_{\  \nu}\nf^{\,\nu}
-\cchi^{\rmc}_{\  \nu}\nc^{\,\nu}\, .\label{88}\fe}
This can be rewritten in the manifestly symmetric standard form 
{\be\TT^{\,\mu\nu}_{\!_{\rm int}}=\rrho_{\rm_{int}}\uu^\mu \uu^\nu
+2\QQ_{\rm_{int}}^{\,(\mu}\uu^{\nu)}+\PP_{\!\rm_{int}}^{\,\mu\nu}
\label{90}\, ,\fe}
with  
{\be  \QQ_{\rm_{int}}^{\,\nu} \uu_\nu=0\, ,\hskip 1 cm
\PP_{\!\rm_{int}}^{\,\mu\nu}\uu_\nu=0\, ,\fe}
in which the symmetric internal pressure tensor 
and internal momentum density vector will be given by
{\be \PP_{\!\rm_{int}}^{\mu\nu}=\PPsi\gamm^{\mu\nu}-2\,\gamm^{\rho\mu}
\gamm^{\sigma\nu}\frac{\partial\LLambda_{\rm_{int}}}
{\partial \ggamma^{\rho\sigma}} \, ,\hskip 1 cm \QQ_{\rm_{int}}^{\,\nu}
=\cchi^{_\Vert}\nn_{^\perp}^{\,\nu}\, ,\label{91}\fe}
while finally the internal mass density $\rrho_{\rm_{int}}$ will be  
given by an expression of the pseudo-Hamiltonian form
{\be  \cc^2\rrho_{\rm_{int}}=\pp^{_\perp}_{\ \nu}\,\nn_{^\perp}^{\,\nu}
-\LLambda_{\rm_{int}}\, .\label{92}\fe}
To put the total into the same standard form
{\be\TT^{\mu\nu}=\rrho\uu^\mu \uu^\nu
+2\QQ^{\,(\mu}\uu^{\nu)}+\PP^{\mu\nu}\, ,\label{93}\fe}
whenever the rest mass parameter $\mm$ is taken to be non-zero
it will be necessary to use the correspondingly augmented quantities
{\be \PP^{\mu\nu}=\frac{\mm}{\nn_\rmf}\,\nn_{^\perp}^\mu\,
\nn_{^\perp}^\nu+\PP_{\!\rm_{int}}^{\mu\nu}\, ,\hskip 1 cm
\QQ^{\,\mu}= \mm\gamm^\mu_{\,\nu}(\nn_{\rmc}^{\,\nu}
+\nn_{\rm c}^{\,\nu})+ \QQ_{\rm_{int}}^{\,\mu} \, ,\label{94}\fe}
so that, finally, we have
{\be\rrho=\mm\left(\nn_\rmc+\frac{\nn_{^\Vert}^{\,2}}{\nn_\rmf}
\right)+\rrho_{\rm_{int}}\, ,\fe}
which can be rewritten as a Legendre type transformation
formula giving the action density in terms of the rest mass
energy density as
 {\be \tilde\LLambda=\left(\mm\gamm[\vv]_\nu+\pp^{_\Vert}_{\ \nu}
\right)\nn_{^\perp}^{\,\nu}-\cc^2\rrho\, .\fe}

\section{Elastic energy contributions}

\subsection{Relaxed contribution}

As in the multiconstituent fluid case \cite{CCH05} it is useful
to decompose the internal action function in the form
{\be \LLambda_{\rm_{int}}=\LLambda_{\rm_{ela}}+\LLambda_{\rm_{ent}}
\label{100} \fe}
so as to obtain a corresponding decomposition
{\be \rrho_{\rm_{int}}= \rrho_{\rm_{ela}}+ \rrho_{\rm_{ent}}\, ,\fe}
in which $\LLambda_{\rm_{ent}}$ and $\rrho_{\rm_{ent}}$
are the parts attributable to entrainment ( which will be dealt
with in Section \ref{entrain}) and $\LLambda_{\rm_{ela}}$ and 
$\rrho_{\rm_{ela}}$ are the static internal contributions that
remain when the relative current contributions $\nn^{\rm_A}$ are 
set to zero. For this static part (but not for the rest) the action 
density will just be the opposite of the elastic energy density $\EU$ 
as defined by setting
{\be  \EU =\rrho_{\rm_{ela}}\cc^2\, , \label{101a}\fe}
which can be seen from (\ref{92}) to correspond to
{\be \LLambda_{\rm_{ela}}=-\EU \, .\label{101b}\fe}

Whenever the comoving constituent can be considered to be a 
sufficiently good electrical conductor -- as will be the case (on a 
macroscopic scale large compared with the scale of the relevant 
plasma frequency) in all the neutron star layers for which the present 
model is intended -- it will be possible, in the manner recently 
described in the purely elastic case \cite{CCC}, to include allowance for 
electromagnetic coupling just in terms of a ``frozen in'' field, with  
antisymmetric material base-space components, $\FF_{\rm_{AB}}$ say, that 
(like the components $\nn_{\rmI_{\rm ABC}}$ of the ionic measure) will be 
fixed functions of the material position coordinates $\qq^{\rm_A}$. In 
such a case there will be a part, $\EU_{\rm_{mag}}$ say, of the elastic 
energy density that depends on the $\qq^{\rm_A}$ not just directly but 
also via its dependence on the components  $\FF_{\rm_{AB}}$, which
themselves depend on the $\qq^{\rm_A}$, and it will commonly be 
helpful to separate this magnetic part out for separate treatment 
(as described in Subsection \ref{magen}) in a decomposition of the form
{\be \EU=\EU_{\rm_{ins}} +\EU_{\rm_{mag}} \, ,\hskip 1 cm
\LLambda_{\rm_{ela}}=\LLambda_{\rm_{ins}} 
+\LLambda_{\rm_{mag}} \, ,\fe}
in which the $\EU_{\rm_{ins}}$ is the strictly static remainder
(the only part that would be needed if the material were an 
electric insulator) that will be left when the frozen components
$\FF_{\rm_{AB}}$ are set to zero.
      
It will usually be convenient to further decompose the strictly static 
(meaning non magnetic) part of the elastic energy contribution in the 
form
{\be \EU_{\rm_{ins}}=\EU_{_\bigcirc}+\EU_{\rm_{sol}}\, ,\hskip 1 cm
\EU_{_\bigcirc}=-\LLambda_{_\bigcirc}
\label{54} \fe}
in which $\EU_{_\bigcirc}$ is the part that remains in a relaxed 
configuration for which $\EU$ is minimised for given values of the 
independent current components $\nn_{\rmc}$, $\nn_\rmf$, $\nn^{\rm_A}$,
and of the determinant of the induced metric $\gamm^{\rm_{AB}}$.  
Fixing this determinant is equivalent to fixing the value of the 
conserved number density $\nn_{\rm I}$ that is specified by (\ref{55}). 
We shall use the notation $\check\gamm{^{\rm_{AB}}}$ and 
$\check\gamm_{\rm_{AB}}$ respectively for the corresponding relaxed 
values of $\gamm^{\rm_{AB}}$ and its inverse $\gamm_{\rm_{AB}}$ (as 
defined by $\gamm_{\rm_{AB}} \gamm^{\rm_{BC}}=\delta^{\rm_C}_{\rm_A}$) 
at which, for given values of $\nn_{\rm I}$, $\nn_{\rmc}$, $\nn_\rmf$, 
the energy minimisation occurs. Thus substitution of 
$\check\gamm{^{\rm_{AB}}}$ for $\gamm^{\rm_{AB}}$ in the solidity 
term $\EU_{\rm_{sol}}$, or in the total $\EU$, will give a generically 
reduced value 
{\be \check \EU_{\rm_{sol}}\leq \EU_{\rm_{sol}}\, ,\hskip
1 cm \check \EU\leq \EU \, ,\label{102a}\fe}
but it will have no effect on the relaxed part, or on the ionic
number density $\nn_{\rm I}$,  for which we simply get
{\be \check\EU_{_\bigcirc}=\EU_{_\bigcirc}\, ,\hskip
1 cm \check \nn_{\rm I} =\nn_{\rm I}\, .\label{102b}\fe}

The relaxed contribution will evidently be of fluid type with a 
generic variation of the form
{\be \delta \EU_{_\bigcirc}=-\delta\LLambda_{_\bigcirc}=
\cchi^{\,\Vert}_{_\bigcirc}\delta\nn_{^\Vert}+\cchi^\rmc_{_\bigcirc}
\delta\nn_\rmc+\cchi^\rms_{_\bigcirc}
\delta\nn_\rmI -\lamb^{\,\rms}_{_{\bigcirc\rm A}}
\delta\qq^{\rm _A}\, ,\label{102c}\fe}
(in which the final term allows for the possibility of built in 
inhomogeneity in addition to the stratification due just to the 
variation of the atomic number ratio $\Armc$) with 
{\be \delta\nn_\rmI=\frac{_1}{^2}\,\nn_\rmI\,\gamm_{\rm_{AB}}
\delta\gamm^{\rm_{AB}} + \frac{\partial\nn_\rmI}
{\partial \qq^{\rm_A}}\delta\qq_{\rm_A}\, .\label{103a}\fe}
The relaxed contribution to the pressure tensor as given by (\ref{91}) 
will have  an expression of the familiar isotropic form
{\be \PP_{\!_\bigcirc}^{\,\rm_{AB}}= \PP_{\!_\bigcirc}\gamm^{\rm_{AB}}
\, ,\hskip 1 cm\PP_{\!_\bigcirc}=\cchi^{\,_\Vert}_{_\bigcirc}
\nn_{^\Vert}+\cchi^{\,\rmI}_{_\bigcirc}\nn_\rmI-\EU_{_\bigcirc}
\, ,\hskip 1 cm \cchi^{\,\rmI}_{_\bigcirc}=\cchi^{\,\rms}_{_\bigcirc}+
\Armc\cchi^{\,\rmc}_{_\bigcirc}\, .\label{104a}\fe}

\subsection{Solidity contribution}

Let us now consider the solidity contribution $\EU_{\rm_{sol}}$
whose job is to allow for the effect (in $\EU_{\rm_{ins}}$) of 
deviations of $\gamm_{\rm_{AB}}$ from its relaxed value 
$\check\gamm_{\rm_{AB}}$ (as determined by the scalars $\nn_\rmf$, 
$\nn_\rmc$, $\nn_\rmI$). Such deviations can conveniently be accounted 
for \cite{CQ72} in terms of the constant volume shear tensor whose 
material base-space representation is specified as
{\be \sigm_{\rm_{AB}}=\frac{_1}{^2}(\gamm_{\rm_{AB}}
-\check\gamm_{\rm_{AB}})\, ,\label{105a}\fe}
which means that the corresponding space time tensor will be 
given by
{\be \sigm_{\mu\nu}= \sigm_{\rm_{AB}}\qq^{\rm_A}_{,\mu}
\qq^{\rm_B}_{,\nu}=\frac{_1}{^2}(\gamm_{\mu\nu}-
\check\gamm_{\mu\nu})\, .\label{105b}\fe}
In most applications to behaviour of a perfectly elastic
(rather than plastic or other more complicated) kind, it will
be sufficient to use an ansatz of quasi-Hookean 
type \cite{CQ72}, meaning one in which the rigidity contribution
has a homogeneously quadratic dependence on the deviation
(\ref{105a}) in the sense that it will be given by an expression
of the form
{\be  \EU_{\rm_{sol}}=\frac{_1}{^2}\,\check{\Shear}{^{\rm_{ABCD}}}
\,\sigm_{\rm_{AB}}\sigm_{\rm_{CD}}
\, ,\label{106}\fe}
with
{\be \check{\Shear}{^{\rm_{ABCD}}}=
\check{\mit\Shear}{^{\rm_{(AB)(CD)}}}=\check{\Shear}{^{\rm_{CDAB}}}
\, \label{06a}\fe}
in which $\check{\Shear}^{\rm_{ABCD}}$ is the relevant solidity
tensor, for which the check symbol is used to indicate that, for
 a given value of the material position coordinates $\qq^{\rm_A}$,
it depends only on the scalars $\nn_\rmf$ and $\nn_\rmc$ and 
(via $\nn_\rmI$) on the relaxed metric $\check\gamm_{\rm_{AB}}$.
The condition that $\gamma_{\rm_{AB}}$ and $\check\gamm_{\rm_{AB}}$ 
must have the same determinant entails that on the 3 dimensional 
material base the symmetric shear tensor $\sigm_{\rm_{AB}}$ will have 
only 5 (instead of 6) independent components, and more specifically 
that to first order it will be trace free with respect to either the 
actual metric $\gamm_{\rm_{AB}}$ or the  relaxed 
$\check\gamm_{\rm_{AB}}$. It is therefore necessary to impose a 
corresponding restriction to completely fix the specification of 
the solidity tensor $\check{\Shear}{^{\rm_{ABCD}}}$, which can be 
conveniently done \cite{CQ72} by requiring that it be trace free 
with respect to the relaxed metric
{\be \check{\Shear}{^{\rm_{ABCD}}\check\gamm_{\rm_{CD}}}=0
\, .\label{06b}\fe}

It is to be remarked  that an alternative procedure that would be 
equivalent for most practical purposes would be to work in an 
analogous manner with a shear tensor constructed as a difference 
between contravariant rather than covariant versions of the actual 
and relaxed forms of the induced metric. Another similarly equivalent 
possibility would be to use a judicious compromise between these 
alternatives, of the kind that has recently been shown \cite{KS02} 
to be technically advantageous for certain purposes. In the small 
shear limit that is physically relevant, the differences between such 
alternatives are only of quadratic order ${\cal O}\{\sigm^2\}$ for 
the shear itself and of higher order for the energy density
$\EU_{\rm_{sol}}$, and will therefore usually be too small to matter 
in practise, so the question of whether to use a prescription of the 
originally proposed kind \cite{CQ72} or of a more sophisticated 
alternative kind \cite{KS02} can be made just on the basis 
of mathematical convenience depending on the context of application. 

The specification of the solidity tensor is not by itself sufficient
to complete the specification of the elastic system, as it is also 
necessary to specify the dependence on $\nn_\rmI$ of the relaxed 
inverse metric $\check\gamm^{\rm_{AB}}$. The simplest possibility
is that of what has been termed a perfect solid \cite{CQ72},
meaning one for which the elastic structure at each material position 
is isotropic with respect to the relaxed metric, which in that case 
can vary only by a conformal factor. This means that it will be given 
in terms of its value $\gamm_0^{\,\rm_{AB}}$ say at some fixed 
reference value $\nn_0$ say of the ionic number density $\nn_\rmI$ by
{\be \check\gamm^{\rm_{AB}} =(\nn_I/\nn_0)^{2/3}\gamm_0^{\,\rm_{AB}}
\, ,\hskip 1 cm\check\gamm_{\rm_{AB}} =(\nn_0/\nn_\rmI)^{2/3}
\gamm_{0\rm_{AB}}\, \label{07}\fe} 

In the case of a solid structure that is isotropic (as will typically
be the case on a macroscopic scale after averaging over randomly oriented
mesoscopic crystals) the solidity tensor in the quasi Hookean ansatz
will simply have to be given in terms of the relevant scalar shear 
modulus $\check\Shear$ by the formula
{\be \check{\Shear}{^{\rm_{ABCD}}}= 2\check\Shear\,(\check\gamm^{\rm_{A(C}}
\check\gamm^{\rm_{D)B}}-\frac{_1}{^3} \check\gamm^{\rm_{AB}}
\check\gamm^{\rm_{CD}})\, ,\label{107a}\fe}
so that (\ref{106}) will give the simple formula
{\be  \LLambda_{\rm_{sol}} = -\EU_{\rm_{sol}}=-\check\Shear\,\sigm^2
\, ,\label{07c}\fe} 
in which the scalar shear magnitude $\sigm$ is defined by the formula
{\be \sigm^2=\check\gamm^{\rm_{AB}}\check\gamm^{\rm_{CD}}\sigm_{\rm_{BC}}
\sigm_{\rm_{DA}}-\frac{_1}{^3}(\check\gamm^{\rm_{AB}}
\sigm_{\rm_{AB}})^2\, ,\label{07e}\fe}
in which the final term will in practise be negligible since of quartic 
order, ${\cal O}\{\sigm^4\}$, in the small $\sigm$ limit that is relevant, 
as the trace is already of quadratic order, 
$\check\gamm^{\rm_{AB}} \sigm_{\rm_{AB}}= {\cal O}\{\sigm^2\}$.
Under these conditions the solidity contribution will provide a pressure
tensor given by the formula
{\be \PP_{\!\rm_{sol}}^{\,\rm_{AB}}= -\check{\Shear}{^{\rm_{ABCD}}}
\sigm_{\rm_{CD}} +\PP_{\!\rm_{sol}}\gamm^{\rm_{AB}}\, ,\label{08} \fe}
of which the final term is a pressure contribution given by
{\be \PP_{\!\rm_{sol}}=\left(\nn_\rmf\frac{\partial\check\Shear}
{\partial\nn_\rmf}+\nn_\rmc\frac{\partial\check\Shear}{\partial\nn_\rmc}
+\nn_\rmI\frac{\partial\check\Shear}{\partial\nn_\rmI}+\frac{\check\Shear}{3}
\right)\sigm^2\, ,\label{08a}\fe}
which, since it is of quadratic order in $\sigm$, will be negligible 
compared with the first (linear order) term for most practical purposes. 
(It is to be noted that the sign of the last of these quadratic order 
terms was one of the errata \cite{CQ77} in the original treatment 
\cite{CQ72}.) The corresponding pressure adjustment contribution for the 
canonical formula (\ref{87})  will be given by
{\be \LP_{\!\rm_{sol}\nu}^{\,\mu} =\gamm_{\nu\lambda}
\check\Shear{^{\lambda\mu\rho\sigma}}\sigm_{\rho\sigma}-\left(
\nn_\rmI\frac{\partial\check\Shear}{\partial\nn_\rmI}+\frac{4\check\Shear}{3}
\right)\sigm^2\gamm^\mu_{\ \nu}\, ,\label{08c}\fe}
in which, again, the quadratic order term at the end will be negligible in 
practise.

\subsection{Magnetic contribution}
\label{magen}

The presence of a frozen in magnetic field (whose effect is likely 
to be significant in applications to ordinary pulsars, and of 
paramount importance in the particular case of magnetars) is 
representable \cite{CCC}  by the specification of a fixed and closed 
material base-space two-form field with components $\FF_{\rm_{AB}}$. 
The closure condition means that it will be locally expressible in 
terms of a fixed gauge 1-form field with components $\AA_{\rm _A}$ 
on the material base manifold as its exterior derivative
{\be \FF_{\rm_{AB}}=2 \AA_{\rm_{[B,A]}}\label{130}\fe}
The corresponding space-time tensor field
{\be \Ff_{\mu\nu}=\FF_{\rm_{AB}}\qq^{\rm_A}_{,\nu}
\qq^{\rm_B}_{,\nu}\fe}
will therefore be given by
{\be \Ff_{\mu\nu}=\Aa_{[\nu,\mu]}\, ,\hskip 1 cm \Aa_\nu=\AA_{\rm_A}
\qq^{\rm_A}_{,\nu}\fe}
and will automatically satisfy the perfect conductivity condition
{\be \Ff_{\mu\nu}\uu^\nu=0 \, ,\fe}
This condition is interpretable as meaning that field has no 
electric part, and so will be entirely determined by its magnetic
part which will have material base components $\BB^{\rm_A}$
given it terms of the alternating tensor $\epsil^{\rm_{ABC}}$
associated with the induced metric $\gamm^{\rm_{AB}}$ by
{\be \BB^{\rm_A}=\frac{_1}{^2}\epsil^{\rm_{ABC}}\FF_{\rm_{BC}}
\, .\fe}

If polarisation effects can be neglected, which will probably be
a good approximation in the crust (though not the core) of a
neutron star, then the magnetic energy contribution will just be
the negative of the usual vacuum action density associated with the
field $\Ff_{\mu\nu}$, which in terms of its materially projected
components gives
{\be \EU_{\rm_{mag}}=\frac{\BB^2}{8\pi}\, ,\hskip 1 cm
\BB^2=\gamma_{\rm_{AB}}\BB^{_A}\BB^{_B}=\frac{_1}{^2}\gamma^{\rm_{AB}}
\gamma^{\rm_{CD}}\FF_{\rm_{AC}}\FF_{\rm_{BD}}\, .\fe}
from which one obtains
{\be \LmP_{\!\rm_{mag}}^{\,\rm_{AB}}=\frac{1}{4\pi}(\BB^{\rm _A}\BB^{\rm _B}
-\BB^2 \gamm^{\rm_{AB}})\, .\fe}
It follows that the magnetic pressure tensor will be given by
{\be  \PP_{\!\rm_{mag}}^{\,\rm_{AB}}=\frac{1}{8\pi}
(\BB^2 \gamm^{\rm_{AB}}-2\BB^{\rm _A}\BB^{\rm _B})\, ,\fe}
so that in terms of the magnetic field covector
{\be \Bb_\mu=\gamm_{\rm_{AB}}\BB^{\rm_B}\qq^{\rm_A}_{,\mu}\, ,\fe}
one obtains an expression of the familiar form,
{\be \TT_{\!\rm_{mag}}^{\,\mu\nu}=\frac{1}{8\pi}\left(\frac{\BB^2}{\cc^2}
\uu^\mu\uu^\nu+\BB^2\gamm^{\mu\nu}-2\BB^\mu\BB^\nu\right)\, ,
\label{138}\fe}
for the magnetic stress energy tensor.
\section{Effect of the current}
\label{entrain}

\subsection{Entrainment contribution}

In general the entrainment action function $\LLambda_{\rm_{ent}}$ will 
depend, for a given values of $\qq^{\rm_A}$, on the relative current
components $\nn^{\rm_A}=\qq^{\rm_A}_{,\nu}\nn_{^\perp}^{\,\nu}$
as well as on the scalar magnitudes $\nn_\rmf$ , $\nn_\rmc$, 
and the induced metric components $\gamm^{\rm_{AB}}$  so its 
generic variation will have the form
{\be \delta\LLambda_{\rm_{ent}}= -\cchi^{\,_\Vert}_{\!\rm_{ent}}
\delta\nn_{^\Vert}- \cchi^{\,\rmc}_{\!\rm_{ent}}\delta\nn_\rmc+
\frac{\partial\LLambda_{\rm_{ent}}}{\partial \gamm^{\rm_{AB}}}
\delta\gamm^{\rm_{AB}}+\frac{\partial\LLambda_{\rm_{ent}}}
{\partial \nn^{\rm_A}}\delta\nn^{\rm_A}+\frac
{\partial\LLambda_{\rm_{ent}}}{\partial \qq^{\rm_A}}\delta\qq^{\rm_A}
\, .\label{64}\fe}

The entrainment action $\LLambda_{\rm_{ent}}$ is characterised by the 
condition that it vanishes when the relative current components 
$\nn^{\rm_A}$ are set to zero, so when these components are sufficiently 
small, as will typically be the case, it will be a good approximation to 
take this contribution to have the homogeneous quadratic form
{\be\LLambda_{\rm_{ent}}=\frac{1}{2\nn_{^\Vert}}\mmm^{^\perp}_{\rm_{AB}}
\nn^{\rm_A}\nn^{\rm_B}\, ,\label{64c}\fe}
in which the entrainment mass tensor has components
$\mmm^{^\perp}_{\rm_{AB}}$ that, like the static action
contribution, are independent of the current components $\nn^{\rm_A}$,
so that the corresponding partial derivative in (\ref{64}) will be 
given by
{\be\pp^{_\perp}_{\,\rm_A}=\frac{\partial\LLambda_{\rm_{ent}}}
{\partial \nn^{\rm_A}}= \frac{1}{\nn_{^\Vert}}\mmm^{^\perp}_{\rm_{AB}}
\nn^{\rm_B}\, .\label{64d}\fe}

It is conceivable that the relaxed action function might involve a 
built in anisotropy favouring relative currents in some particular 
direction, but in cases of the simplest kind, to which the remainder 
of this subsection and the next will be restricted, this function
$\LLambda_{\rm_{ent}}$ will be of purely fluid type in the sense that 
for given values of $\qq^{\rm_A}$ it will  depend only on the set of 
four scalar magnitudes consisting of $\nn_\rmI$, $\nc$, $\nf$ together 
with the relative current magnitude $\nn_{^\perp}$ that is defined in 
terms of the (unrelaxed) metric value $\gamm_{\rm_{AB}}$ which -- 
using the material index lowering operation specified by the induced
 metric $\gamm^{\rm_{AB}}$ -- will be given by
{\be \nn_{^\perp}^{\,2} =\nn^{\rm_A}\nn_{\rm_A}\, ,\hskip 1 cm 
\nn_{\rm_A}=\gamm_{\rm_{AB}}\nn^{\rm_B}\, ,\hskip 1 cm \gamm_{\rm_{AB}}
\gamm^{\rm_{BC}}=\delta^{\rm_C}_{\rm_A}\, .\label{110}\fe}
Such a functional dependence provides an expansion
{\be \delta\LLambda_{\rm_{ent}}=-\cchi^\rmf_{\rm_{ent}}\delta\nn_\rmf-
\cchi^\rmc_{\rm_{ent}}\delta\nn_\rmc-\cchi^\rms_{\rm_{ent}}\delta\nn_\rmI 
+\lamb^{\,\rms}_{\!\rm_{ent\, A}}\delta\qq^{\rm _A}+\frac
{\partial\LLambda_{\rm_{ent}}}{\partial\nn{_{^\perp}^{\,2}}}
\delta\nn{_{^\perp}^{\,2}}\, ,\label{111}\fe}
of similar form to the perfect fluid contribution (\ref{102c}) 
but with an extra term involving a partial derivative that
provides an expression of the form (\ref{64d}) in terms
an isotropic mass tensor given by
{\be \mmm^{^\perp}_{\rm_{AB}}=\mmm^\rmf_{\,\rmc}\gamm_{\rm_{AB}}\, ,
\hskip 1 cm \mmm^\rmf_{\,\rmc}= 2\nn_\rmf\frac 
{\partial\LLambda_{\rm_{ent}}}{\partial\nn{_{^\perp}^{\,2}}}
\, ,\label{112}\fe}
while the other partial derivatives in (\ref{64}) will be given by
{\be \frac{\partial\LLambda_{\rm_{ent}}}{\partial\gamm^{\rm_{AB}}}
=-\frac{_1}{^2}\left(\frac{\mmm^\rmf_{\,\rmc}}{\nn_{^\Vert}}\,
\nn_{\rm_A}\nn_{\rm_B} +{\cchi^{\,\rms}_{\!\rm_{ent}}}\,
\nn_\rmI\,\gamma_{\rm_{AB}}\right) \, ,\hskip 1 cm
\frac{\partial\LLambda_{\rm_{ent}}}{\partial \qq^{\rm _A}}
=\lamb^{\rms}_{\!\rm_{ent\,A}}-\cchi^{\,\rms}_{\!\rm_{ent}}
\frac{\partial\nn_\rmI}{\partial\qq^{\rm_A}}\, . \label{113}\fe}
The scalar $\mmm^\rmf_{\,\rmc}$ introduced in this way is identifiable 
as the increment $\mmm^\rmf_{\,\rmc}=\mmm_\star-\mm$ of the effective 
mass $\mmm_\star$ of the free baryons (meaning the superfluid neutrons) 
as compared with the ordinary baryonic mass $\mm$. This mass increment 
is expected to be positive (and in some layers large) \cite {CCH05} in 
the solid neutron star crust, but (moderately) negative in the fluid 
layers below.

\subsection{Relaxed action contribution}
\label{relac}

It will be useful for many purposes to replace the decomposition
(\ref{100}) of the internal action density by an alternative
decomposition of the form
{\be\LLambda_{\rm_{int}}=\LLambda_{\rm_{lax}}+\LLambda_{\rm_{rig}}
\, ,\hskip 1 cm \LLambda_{\rm_{rig}}=\LLambda_{\rm_{sol}}
+\LLambda_{\rm_{mag}}\, , \label{120}\fe}
in which the relaxed -- meaning shear independent -- part will 
evidently consist of the combination 
 {\be\LLambda_{\rm_{lax}}=\LLambda_{_\bigcirc}+\LLambda_{\rm_{ent}}
\, .\label{121}\fe}

The use of such a combination is particularly convenient whenever the 
entrainment contribution is of the isotropic type characterised by 
the variation expansion (\ref{111}), in which case  the complete 
relaxed action density will have a variation given by an expansion of 
the analogous form
{\be \delta\LLambda_{\rm_{lax}}=-\cchi^\rmf_{\rm_{lax}}\delta\nn_\rmf-
\cchi^\rmc_{\rm_{lax}}\delta\nn_\rmc-\cchi^\rms_{\rm_{lax}}\delta\nn_\rmI 
+\lamb^{\,\rms}_{\rm_{lax\, A}}\delta\qq^{\rm _A}
+\frac{\mmm^\rmf_{\ \rmc}}{2\nn_{^\Vert}}\,
\delta\nn{_{^\perp}^{\,2}}\, ,\label{122}\fe}
with
{\be \cchi^\rmf_{\rm_{lax}}=\cchi^\rmf_{_\bigcirc}\!+\!\cchi^\rmf_{\rm_{ent}}
\, ,\ \ \  \cchi^\rmc_{\rm_{lax}}=\cchi^\rmc_{_\bigcirc}\!
+\!\cchi^\rmc_{\rm_{ent}}\, ,\ \ \ \cchi^\rms_{\rm_{lax}}=
\cchi^\rms_{_\bigcirc}\!+\!\cchi^\rms_{\rm_{ent}}\, ,\ \ \
\lamb^\rms_{\rm_{lax\, A}}=\lamb^\rms_{_{\bigcirc A}}\!+\!
\lamb^\rms_{\rm_{ent\, A}}\, .\label{123}\fe}

It can be seen that the relaxed contribution to the ``live'' action 
variation (\ref{83}) will simplify to provide an expression of the form
{\be \delta^{_\heartsuit\!}\LLambda_{\rm_{lax}}=
 \cchi^{\,\rmf}_{\!_{\rm lax}\nu} \, \delta\nn_{\rm f}^{\,\nu}+ 
\cchi^{\,\rmc}_{\!_{\rm lax}\nu}\, \delta\nn_{\rm c}^{\,\nu} 
-\cchi^{\rms}\delta^{_\heartsuit\!}\nn_{\rmI}
+\lamb^{\,\rms}_{\rm_{lax\, A}}
\delta\qq^{\rm _A}\, ,\label{60}\fe}
in which  $\cchi^{\,\rmf}_{\!_{\rm lax}\nu}$ and 
$\cchi^{\,\rmc}_{\!_{\rm lax}\nu}$ can be read out as the
relaxed parts of the internal momenta given by (\ref{85}).
Since we can write $ \cc^2\delta^{_\heartsuit\!}\nn_{\rm I}=-\uu_\nu\,
\delta^{_\heartsuit\!}\nn_{\rm I}^{\, \nu}$, it can be seen that, 
as the analogue of these free and confined particle 4-momentum 
contributions, we shall also be able to read out a corresponding
ionic stratification 4-momentum contribution  
$\cchi^{\,\rms}_{\!_{\rm lax}\nu}$, so as to obtain a complete set
of relaxed internal momentum covectors that will be 
expressible in terms of the relative flow velocity vector
introduced in (\ref{53a}) by
{\be \cchi^{\,\rmf}_{\!_{\rm lax}\nu}=\frac{\cchi^{_\Vert}_{\!_{\rm lax}}}
{\cc^2}\,\uu_\nu+\mmm^\rmf_{\ \rmc}\, [\vv]_\nu\, ,\label{61a}\fe}
{\be \cchi^{\,\rmc}_{\!_{\rm lax}\nu}=\frac{\cchi^{\,\rmc}_{\!_{\rm lax}}}
{\cc^2}\uu_\nu+\frac{\nn_{^\Vert}}{\nn_\rmc}
\left(\frac{\cchi^{_\Vert}_{\!\rm_{lax}}}{\cc^2}
-\mmm^\rmf_{\ \rmc}\right)\, [\vv]_\nu\, ,\label{61b}\fe}
{\be\cchi^{\,\rms}_{\!_{\rm lax}\nu}=\frac{\cchi^{\,\rms}_{\!_{\rm lax}}}
{\cc^2}\uu_\nu\, ,\label{61c}\fe}
It follows that the corresponding contribution to the generalised pressure 
scalar (\ref{88}) will be given by
{\be \PPsi_{\rm_{lax}}=\LLambda_{\rm_{lax}}+\nn_{^\Vert}\cchi^{\,_\Vert}
_{\!\rm_{lax}}+\nn_\rmc\cchi^{\,\rmc}_{\!\rm_{lax}}-\mmm^\rmf_{\,\rmc}
[\vv]^2\, .\fe}
In the formula (\ref{84}) for the extra stress, it transpires
that the contributions involving the mass increment 
$\mmm^\rmf_{\ \rmc}$  (proportional to $\partial
\LLambda_{\rm_{lax}}/\partial\nn_{^\perp}^{\,2}$) will cancel out, 
leaving just the part due to stratification which
will be given simply by
{\be \LP^{\, \nu}_{\!\rm_{lax}\,\mu}=-\cchi^{\rms}_{\!_{\rm lax}}
\,\nn_\rmI\, \gamm^\nu_{\ \mu}\, .\fe}
It is to be observed that the need to include this extra term in the
canonical stress energy formula can be avoided by treating the
stratification momentum (\ref{61c}) as if it were on the same
footing as the others, which means including a corresponding
extra term in a supplemented pressure scalar $\PPsi^\rms$ defined by
{\be \PPsi^\rms=\LLambda_{\!\rm_{lax}}-\cchi^{\,\rmf}_{\!\rm_{lax}\nu}
 \nn_\rmf^{\,\nu}-\cchi^{\,\rmc}_{\!\rm_{lax}\nu}\nn_\rmc^{\,\nu}
-\cchi^{\,\rms}_{\!\rm_{lax}\nu}\nn_\rmI^{\,\nu}=\PPsi_{\rm_{lax}}+
\cchi^{\,\rms}_{\!\rm_{lax}}\nn_\rmI\, .\fe} 

Combining  the relaxed internal contribution $\LLambda_{\rm_{lax}}$ with 
the ballistic contribution $\tilde\LLambda_{\rm_{bal}}$ one obtains the 
complete liquid (meaning hydrodynamic as opposed to solid or 
magnetohydrodynamic) part of the action density in the form
{\be \tilde\LLambda_{\rm_{liq}}=\tilde\LLambda_{\rm_{bal}}+
\LLambda_{\rm_{lax}}=\tilde\LLambda-\LLambda_{\rm_{rig}}\, ,\fe}
for which the corresponding liquid contributions to the momentum
 covectors  (\ref{51}) will be given (in terms of the relative flow
velocity $[\vv]^\nu=\nn_{^\Vert}^{-1}\nn_{^\perp}^{\,\nu}$) by
{\be \mmu^{\,\rmf}_{\!\rm_{liq}\nu}=\mm \,\gamm(\uu_{\nu}+[\vv]_\nu)
+\cchi^\rmf_{\rm_{lax}\nu}\, ,\hskip 1 cm
\mmu^{\,\rmc}_{\rm_{liq}\nu}=\mm \,\uu_\nu+\cchi^\rmc_{\rm_{lax}\nu}
\hskip 1 cm \mmu^\rms_{\,\nu}=\cchi^\rms_{\rm_{lax}\nu}
\, .\fe}
In terms of these quantities the corresponding liquid contribution 
to the stress energy tensor will be given simply by
{\be \TT^{\,\mu}_{\!\rm_{liq}\nu}= \mmu^{\,\rmf}_{\!\rm_{liq}\nu}
\nn_\rmf^{\,\mu}+\mmu^{\,\rmc}_{\!\rm_{liq}\nu}
\nn_\rmc^{\,\mu}+\mmu^{\,\rms}_{\!\rm_{liq}\nu}
\nn_\rmc^{\,\mu}+\PPsi^\rms\,\ddelta^\mu_{\,\nu}\, ,\fe}
and the corresponding elastically relaxed force contributions acting on free
and confined currents will be given by an ansatz of the standard form
(\ref{(22)}) which gives
{\be \ff^{\rm f}_{\!_{\rm liq} \nu}= 2\nn_{\rm f}^{\,\mu} 
\nabla_{\![\mu}\mmu^{\rm f}\!_{\!_{\rm liq}\nu]}+
\mmu^{\rm f}_{\!_{\rm liq} \nu} 
\nabla_{\!\mu}\nn_{\rm f}^{\,\mu} \, ,\label{70}\fe} 
and 
{\be \ff^{\rm c}_{\!_{\rm liq} \nu}= 2\nn_{\rm c}^{\,\mu} 
\nabla_{\![\mu}\mmu^{\rm c}_{\!_{\rm liq}\nu]}+
\mmu^{\rm c}_{\!_{\rm liq} \nu} \nabla_{\!\mu}\nn_{\rm c}^{\,\mu} 
\, ,\label{71}\fe}
while for the analogously defined force contribution (\ref{(23)}) due to 
stratification acting on the underlying ionic lattice, it can be seen that
similar reasoning leads to an expression of the slightly different form
{\be \ff^\rms_{\!_{\rm liq} \nu}= 2\nn_{\rmI}^{\,\mu} 
\nabla_{\![\mu}\mmu^{\rms}_{\!_{\rm liq}\nu]}+
\lamb^{\,\rms}_{\!\rm_{lax}\nu}\, ,\hskip 1 cm
\lamb^{\,\rms}_{\!\rm_{lax}\nu}=
\lamb^{\,\rms}_{\!\rm_{lax\,A}}\qq^{\rm_A}_{\, ,\nu}\, .\label{72}\fe}

As remarked at the outset, because the confined particles and the ionic 
medium are constrained to move together the confined and supplementary force
densities are not separately meaningful: all that is relevant for
the equations of motion is their sum, the amalgamated ionic force density
(\ref{12}). Its liquid contribution can conveniently be expressed
by introducing the corresponding amalgamated ionic 4-momentum  
that is definable as
 {\be \mmu^{\rmI}\!_{\!_{\rm liq}\nu}= \mmu^{\rms}\!_{\!_{\rm liq}\nu} 
+\Armc\, \mmu^{\rmc}\!_{\!_{\rm liq}\nu}\, ,\label{67}\fe}
(where $\Armc=\nn_\rmc/\nn_\rmI$ is the atomic number as introduced
above). In terms of the amalgamated ionic 4-momentum it can be seen that one 
will simply obtain the formula
{\be \ff^\rmI_{\!_{\rm liq} \nu}= 2\nn_{\rm I}^{\,\mu} \nabla_{\![\mu}
\mmu^{\rmI}_{\!_{\rm liq}\nu]}+\nn_\rmI^{\,\mu}\mmu^{\rm c}_{\!_{\rm liq}\mu}
\nabla_{\!\nu}\Armc+\lamb^{\,\rms}_{\!\rm_{lax}\nu}
\, , \fe}
while the associated contribution to the stress energy tensor 
(\ref{(21)}) will be expressible by the neatly concise formula
{\be \TT^{\ \mu}_{\!_{\rm liq} \nu}=\nn_{\rmf}^{\,\mu} 
\mmu^{\rmf}_{\!_{\rm liq}  \nu}+\nn_{\rmI}^{\,\mu} 
\mmu^{\rmI}_{\!_{\rm liq} \nu}+\PPsi^\rms\ddelta^\mu_{\ \nu}
\, ,\label{75}\fe}
in which we shall have
{\be \PPsi^\rms=\LLambda_{_{\rm lax}}-\nn_{\rmf}^{\,\rho}
\ppi^{\rmf}_{\!_{\rm lax} \rho}-\nn_{\rmI}^{\,\rho}
\ppi^{\rmI}_{\!_{\rm lax} \rho}\, .\fe}

\subsection{Complete description for perfect conducting solid.}

Let us conclude by putting all the pieces together for the case of what 
we refer to as a perfect conducting solid -- meaning a model of the 
qualitatively simplest kind  whose structure is fully isotropic with 
respect to the relaxed metric. In so far as the liquid part is concerned, 
this condition means not only that the incremental mass matrix
has to be isotropic,  as supposed in the preceding subsection, but also
that the extra stratification vector in (\ref{72}) must vanish,
$\lamb^{\,\rms}_{\!\rm_{lax}\nu} =0$.

Replacing the relaxed contribution (\ref{67}) in (\ref{75}) by the 
corresponding total ionic 4-momentum covector given (in terms of the 
confined atomic number $\Armc$)  by
{\be \mmu^{\rmI}_{\,\nu}= \mmu^{\rms}_{\,\nu}+\Armc
 \mmu^{\rmc}_{\,\nu}\, , \label{67a}\fe}
we can apply an analogous tidying up operation to the complete
stress energy tensor (\ref{(21)}) which will thereby acquire the form
{\be \TT^{\mu}_{\ \nu}=(\LLambda-
\nn_{\rmf}^{\,\rho}\mmu^{\rmf}_{\,\rho}-\nn_{\rmI}^{\,\rho}
\mmu^{\rmI}_{\, \rho})\delta^\mu_{\ \nu}+\nn_{\rmf}^{\,\mu} 
\mmu^{\rmf}_{\, \nu}+\nn_{\rmI}^{\,\mu} 
\mmu^{\rmI}_{\, \nu}+\TT_{\!\rm_{mag}\nu}^{\,\mu}
-\LP^{\, \mu}_{\!\rm_{sol}\,\nu}\
\, ,\label{115}\fe}
in which the magnetic contribution is given by (\ref{138}) and the only 
manifest allowance for the effects of solid rigidity is in the final term. 

If we make the further realistic (and chemically invariant) supposition
that (since it arises just from Coulomb interactions between the
confined protons) the shear modulus depends only on $\nn_\rmc$ and
$\nn_\rmI$, then the solidity term will not contribute to the free
particle momentum so we shall have
{\be \mmu^\rmf_{\,\nu}=\mmu^{\,\rmf}_{\!\rm_{liq}\nu}\, ,\hskip
1 cm \ff^\rmf_{\,\nu}=\ff^{\,\rmf}_{\!\rm_{liq}\nu}\, ,\fe}
while the force density acting on the part that is comoving with the
ions will be given by
{\be\ff^\rmI_{\,\nu}=\ff^{\,\rmI}_{\!\rm_{liq}\nu}+
\ff^	{\,\rmI}_{\!\rm_{rig}\nu}
\, ,\fe}
where the rigidity term (allowing for the combined effect of ionic 
solidity and the magnetic field that may be frozen into the 
accompanying degenerate electron plasma) will be given by
{\be \ff^{\,\rmI}_{\!\rm_{rig}\nu}= \nabla_{\!\mu}
\TT^{\,\mu}_{\!\rm_{rig}\nu}\, , \hskip 1 cm
\TT^{\,\mu}_{\!\rm_{rig}\nu}=\TT^{\,\mu}_{\!\rm_{mag}\nu}+
\TT^{\,\mu}_{\!\rm_{sol}\nu}\, .\label{117}\fe}
To linear order in the shear amplitude -- which will be sufficiently
accurate for most practical purposes -- the solidity contribution will 
be given by the expression
{\be \TT^{\,\mu}_{\!\rm_{sol}\nu}=-\LP^{\, \mu}_{\!\rm_{sol}\,\nu}\
+{\cal O}\{\sigm^2\}\, ,\label{116}\fe}
with $\LP^{\, \mu}_{\!\rm_{sol}\,\nu}$ given by the perfect solid
formula (\ref{08c}), which will itself be expressible to first order 
in the shear amplitude $\sigm$ by a prescription of the familiar form
{\be \LP^{\, \mu}_{\!\rm_{sol}\,\nu}=2\check\Shear\,
\sigm^\mu_{\ \nu}+{\cal O}\{\sigm^2\}\, ,\label{118}\fe}
in which $\check\Shear$ is the ordinary shear modulus (which elsewhere is 
commonly denoted by the symbol $\mu$ that, in the present context, 
has already been used for the designation of the material momentum 
components). This means that for practical purposes, for a conducting
solid of the perfect (meaning intrinsically isotropic) type the 
term that is needed \cite{Cham06} to allow for deviation from 
behaviour of (multiconstituent) fluid type will be entirely contained 
in the magnetic contribution of the usual magnetohydrodynamic form 
given by (\ref{138}) and the extra solidity term as given by (\ref{118}), 
on the understanding that $\check\Shear$ is prescribed as a function 
just of $\nn_\rmc$ and $\nn_\rmI$.


\begin{thebibliography}{99}

\bibitem{C89} B. Carter,
``Covariant Theory of Conductivity in Ideal Fluid or Solid Media",
B. Carter, in {\it Relativistic Fluid Dynamics (C.I.M.E., Noto, May 1987)}
ed.  A.M. Anile, \& Y. Choquet-Bruhat, Lecture Notes in Mathematics 
{\bf 1385} (Springer - Verlag, Heidelberg, 1989) 1-64.

\bibitem{M78} G.A. Maugin,
``On the covariant equations of the relativistic electrodynamics
of continua: III elastic solids'',
{\it J. Math. Phys.} {\bf 19} (1978) 1212-1219.

\bibitem{C80} B. Carter, 
``Rheometric structure theory, convective differentiation, and
continuum electrodynamics'',
{\it Proc. Roy. Soc. Lond.} {\bf A372} (1980) 169-200.

\bibitem{IETAL05} G. L. {Israel}, T. {Belloni}, L. {Stella}, Y. {Rephaeli}, D. E. {Gruber},
P. {Casella}, S. {Dall'Osso}, N. {Rea}, M. {Persic}, R. E. {Rothschild},
``The Discovery of Rapid X-Ray Oscillations in the Tail of the SGR 1806-20
Hyperflare'',
{\it Astroph. J.} {\bf 628} (2005) L53-L56.
[astro-ph/0505255]

\bibitem{WS06} A. L. {Watts}, T. E. {Strohmayer}, 
``Detection with RHESSI of High-Frequency
 X-Ray Oscillations in the Tail of the 2004 Hyperflare from SGR 1806-20'',
{\it Astroph. J.} {\bf 637} (2006) L117-L120.
[astro-ph/0512630]

\bibitem{SW05} T. E. {Strohmayer}, A. L. {Watts}, 
``Discovery of Fast X-Ray Oscillations during
the 1998 Giant Flare from SGR 1900+14'',
{\it Astroph. J.} {\bf 632}  (2005) L111-L114.
[astro-ph/0508206]

\bibitem{D98}  R. C. {Duncan}, 
``Global Seismic Oscillations in Soft Gamma Repeaters'',
{\it Astroph. J.} {\bf 498} (1998) L45-L49. 
[astro-ph/9803060]

\bibitem{GSA06} K. Glampedakis, L. Samuelsson, N. Andersson 
 ``Elastic or magnetic? {A} toy
  model for global magnetar oscillations with implications for {Q}{P}{O}s
  during flares'', accepted in {\it Mon. Not. R. Astr. Soc.} (2006) [astro-ph/0605461]

\bibitem{CK92} B. Carter, I.M. Khalatnikov,
``Equivalence of convective and potential variational derivations
of covariant superfluid dynamics'',
{\it Phys. Rev.} {\bf D45} (1992) 4536-4544.

\bibitem{Langlois98} D. Langlois, D. Sedrakian, B. Carter,
``Differential rotation of relativistic superfluid in neutron stars'',
{\it Mon. Not. R. Astr. Soc.} {\bf 297} (1998) 1198-1201.
[astro-ph/9711042]

\bibitem{ACL02} N. Andersson, G.L. Comer, D. Langlois,
"Oscillations of general relativistic superfluid neutron stars",
{\it Phys. Rev.} {\bf D66} (2002) 104002 [gr-qc/0203039] 

\bibitem{Comer04} G.L. Comer,
"Slowly rotating general relativistic superfluid neutron stars with
relativistic entrainment",
{\it Phys. Rev.} {\bf D69} (2004) 123009 [gr-qc/0402015]  

\bibitem{PNC05} R. Prix, J. Novak , G.L. Comer,
"Relativistic numerical models for stationary superfluid neutron stars",
{\it Phys. Rev.} {\bf D71} (2005) 043005 [gr-qc/0410023] 

\bibitem{CQ72} B. Carter, H. Quintana,
``Foundations of general relativistic high-pressure elasticity
theory'',
{\it Proc. Roy. Soc. Lond.} {\bf A331} (1972) 57-83.

\bibitem{C73} B. Carter, 
``Elastic perturbation theory in General Relativity 
and a variational Principle for a rotating solid Star'', 
{\it Commun. Math. Phys.} {\bf 30} (1973) 261-286.

\bibitem{FS75} J.L. Friedman, B.F. Schutz,
``On the stability of relativistic systems'',
{\it Astroph. J.} {\bf 200} (1975) 204-220.

\bibitem{CQ77} B. Carter, H. Quintana,
 ``Gravitational And Acoustic Waves In An Elastic Medium'',
{\it Phys. Rev.} {\bf D16} (1977) 2928-2938. 
 
\bibitem{CQ05}  B. Carter, H. Quintana,
``Stationary Elastic Rotational Deformation of a Relativistic Neutron
Star Model", {\it Astrophys. J.}
{\bf 202}, (1975) 511-522.

\bibitem{Q76} H. Quintana,
``The structure equations of a slowly rotating, fully relativistic 
solid star'',
{\it Astroph. J.} {\bf 207} (1976) 279-288.

\bibitem{ST83} B.L. Schumaker, K.S. Thorne,
``Torsional oscillations of neutron stars'',
{\it Mon. Not. R. Astr. Soc.} {\bf 203} (1983) 457-489.

\bibitem{F90} L.S. Finn,
``Non-radial pulsations of neutron stars with a crust'',
{\it Mon. Not. R. Astr. Soc.} {\bf 245} (1990) 82-91.

\bibitem{P92} D. Priou, 
``The perturbations of a fully general relativistic and rapidly rotating 
neutron star. I - Equations of motion for the solid crust'',
{\it Mon. Not. R. Astr. Soc.} {\bf 254} (1992) 435-452.

\bibitem{KS02} M. Karlovini, L. Samuelsson,``Elastic stars in general 
relativity: I. Foundations and equilibrium models'',
{\it Class. Quantum. Grav.} {\bf 20} (2003) 3613-3648.
[gr-qc/0211026] 
 
\bibitem{KS03} M. Karlovini, L. Samuelsson,
``Elastic stars in general relativity: {I}{I}. Radial perturbations''
{\it  Class. Quant. Grav.} {\bf 21} (2004) 1559-1581.
[gr-qc/0309056]

\bibitem{Beig02} R. Beig, G. Schmidt, ``Relativistic elasticity'',
{\it Class. Quant. Grav.} {\bf 20} (2003) 889-904.
[gr-qc/0211054]

\bibitem{KS04} M. Karlovini, L. Samuelsson,
 ``Elastic stars in general relativity: {I}{I}{I}. Stiff ultrarigid exact 
solutions'', 
{\it Class. Quant. Grav.} {\bf 21} (2004) 4531-4548.
[gr-qc/0401115]

\bibitem{Cham06} N. Chamel, B. Carter, 
``Effect of entrainment on stress and pulsar glitches in stratified
neutron star crust'',
{\it Mon. Not. R. Astr. Soc.} {\bf 368} (2006) 796-803.
[astro-ph/0503044]

\bibitem{CC06} B. Carter, E. Chachoua,
``Newtonian mechanics of neutron superfluid in elastic star crust'',
to be published in {\it Int. J. Mod. Phys.} {\bf D} (2006). [astro-ph/0601658]

\bibitem{CCC} B. Carter, E. Chachoua, N. Chamel,
``Covariant Newtonian and Relativistic dynamics of (magneto)-elastic solid
model for neutron star crust.'',
{\it Gen. Rel. Grav.} {\bf38} (2006) 83-119.
[gr-qc/0507006]

\bibitem{C06} B. Carter,
``Poly-essential and general Hyperelastic World (brane) models'',
contrib. to {\it Micro and Macro Structures of Spacetime}
(Peyresq 2005). [hep-th/0604157]

\bibitem{CCH05} B. Carter, N. Chamel, P. Haensel,
``Entrainment coefficient and effective mass for conduction neutrons
in neutron star crust: {I}{I} macroscopic treatment'',
{\it Int. J. Mod. Phys.} {\bf D} (2006).
[astro-ph/0408083 v3]



\end{thebibliography}
\end{document}